\documentclass{iaus}
\usepackage{graphicx}

\title[Dark Matter in Galaxies]{Where Does The Dark Matter Begin?}

\author[C.S. Kochanek]%
{C.S. Kochanek}

\affiliation{Department of Astronomy, The Ohio State University}

\pubyear{2004}
\volume{225}
\pagerange{1--8}
\date{?? and in revised form ??}
\jname{Impact of Gravitational Lensing on Cosmology}
\editors{Mellier, Y. \& Meylan,G. eds.}
\begin{document}

\maketitle

\begin{abstract}
While there is convincing evidence that the central regions ($r \ll R_e$) 
of early-type galaxies are dominated by stars and that the outer regions 
($r \gg R_e $) are dominated by dark matter, the structure of early-type
galaxies in the transition region (a few effective radii $R_e$) 
between the stars and the dark matter is unclear both locally and in gravitational lenses.  
Understanding the structure of galaxies in this transition region is 
a prerequisite for understanding dark matter halos and how they
relate to the luminous galaxy.  Potentially the best probe of 
this region is the sample of $\sim 80$ strong gravitational lenses.
I review the determination of 
mass distributions using gravitational lenses using image positions,
statistics, stellar dynamics, time delays and microlensing.  While the
present situation is confusing, there is little doubt that the existing
problems can be resolved by further observations.
\end{abstract}

\def\rbar{\langle R\rangle }
\def\dr{\Delta R}
\def\gax{\mathrel{\raise.3ex\hbox{$>$}\mkern-14mu\lower0.6ex\hbox{$\sim$}}}
\def\lax{\mathrel{\raise.3ex\hbox{$<$}\mkern-14mu\lower0.6ex\hbox{$\sim$}}}
\def\gtorder{\mathrel{\raise.3ex\hbox{$>$}\mkern-14mu
             \lower0.6ex\hbox{$\sim$}}}
\def\ltorder{\mathrel{\raise.3ex\hbox{$<$}\mkern-14mu
             \lower0.6ex\hbox{$\sim$}}}
\def\hunits{km~s$^{-1}$~Mpc$^{-1}$}

\firstsection % if your document starts with a section,
\section{Introduction}

\begin{figure}[t]
\begin{center}
\centerline{\includegraphics[width=2.5in]{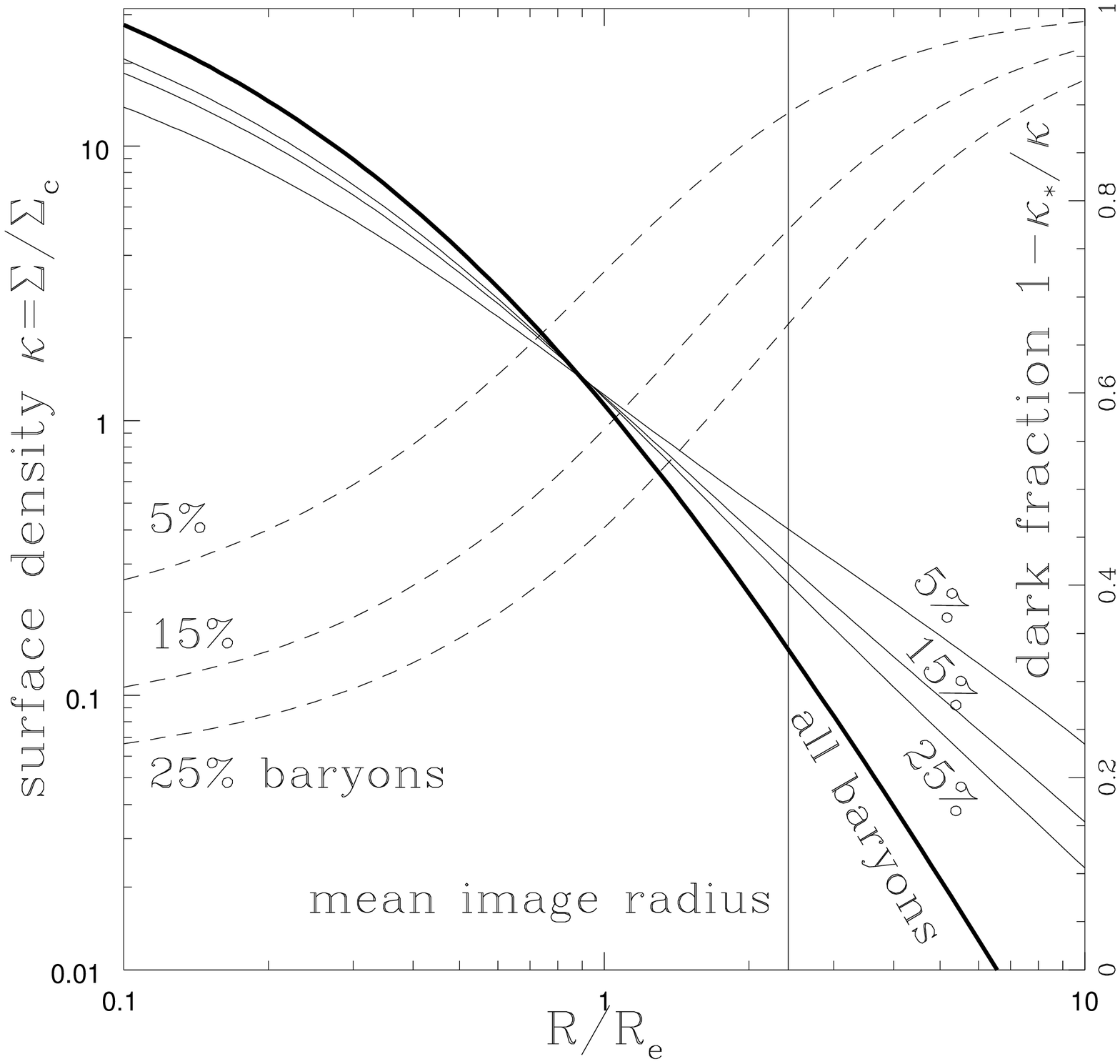}
            \includegraphics[width=2.5in]{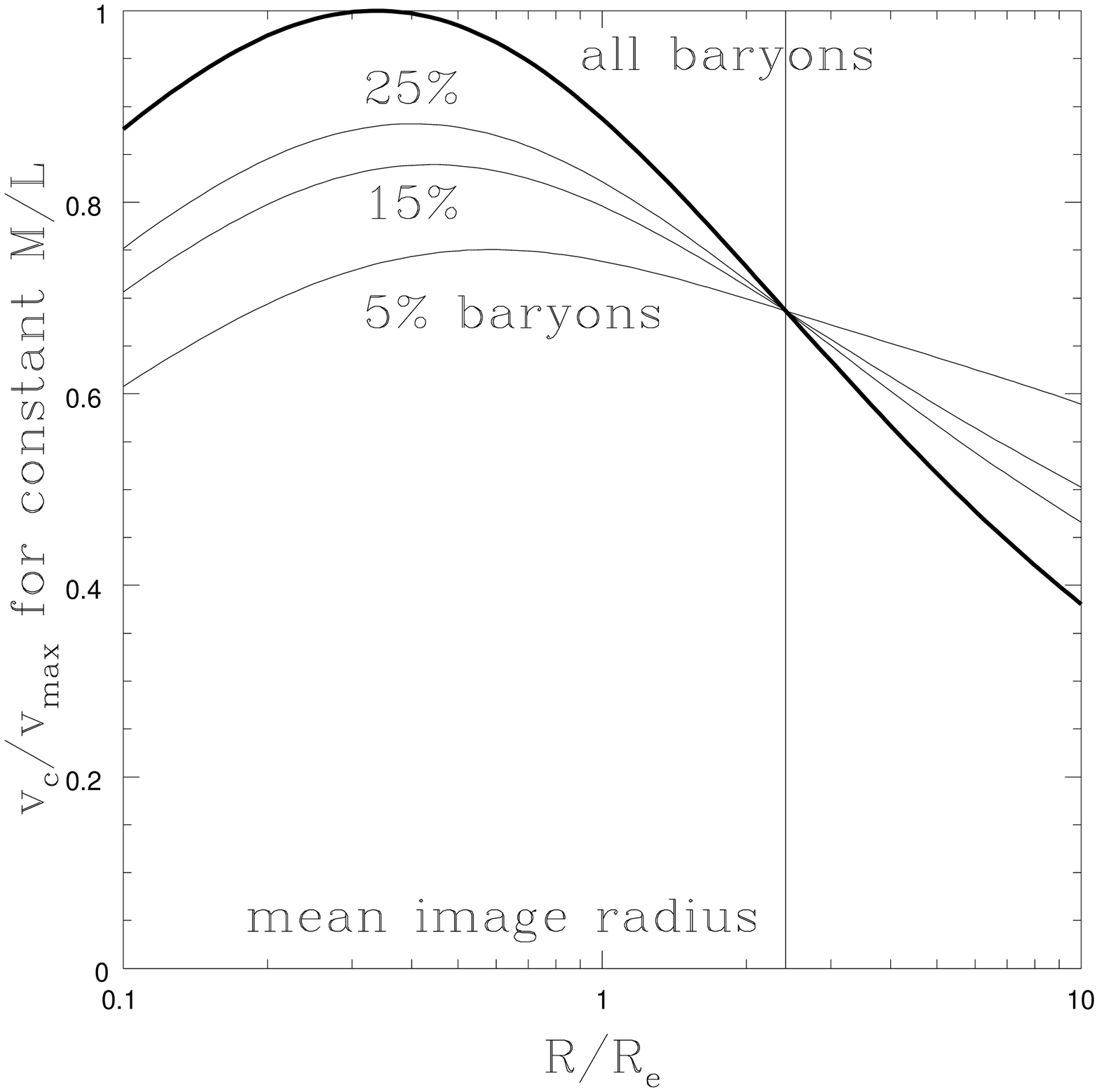}
    }
\end{center}
\caption{ The projected surface density (left) and rotation curve (right) 
   of a lens galaxy.  The mass interior to the mean 
   image radius is determined by the image geometry, the stellar distribution is
   measured with HST, and the dark matter distribution is an adiabatically
   compressed NFW profile.  A constant $M/L$ model with no dark matter is
   shown by the heavy solid line, and the light solid lines show profiles
   where the baryons (stars) represent 25\%, 15\% or 5\% of the total halo
   mass.  The dashed lines (right hand scale) show the fraction of the surface density
   in dark matter, $1-\kappa_*/\kappa$, as a function of radius.  We expect a baryon 
   fraction of $\simeq 16\%$ if all baryons in the halo cool and form stars.
   The logarithmic radial scale greatly exaggerates the apparent decline of
   the rotation curves.
   }
  \label{fig:halo}
\end{figure}

In our current paradigm for the structure of galaxies, the stars we observe
in an early type galaxy should be embedded in a dark matter halo.  
The stars and any associated cold gas can comprise up to 16\% of the halo mass
(e.g. Spergel et al.~\cite{Spergel2003p175}),
although the actual mass fraction could be lower if star formation processes
either eject gas from the halo or maintain it at the virial temperature
of the halo.  The halo structure should be similar to an NFW model (Navarro
et al.~\cite{Navarro1996p563}), although 
the central density of the dark matter may be increased by adiabatic
compression (Blumenthal et al.~\cite{Blumenthal1986p27}).  Unfortunately, 
for early-type galaxies, there is no simple
conserved quantity that can be used to estimate a scale length for the 
stellar component in the way that angular momentum conservation can be
used to estimate it for a disk galaxy.  Fig.~\ref{fig:halo} shows the
expected surface density for a typical lens in terms of the dimensionless 
convergence $\kappa(R)=\Sigma(R)/\Sigma_c$ relevant to gravitational lensing.  

The effective radius $R_e$ of an early-type galaxy provides a characteristic
scale for the luminous parts of the galaxy.
On small scales ($R/R_e < 1$) the galaxy is dominated by the baryonic mass, with
the dark matter contributing only $\sim 20\%$ of the projected density.  This is 
borne out by dynamical studies of the central regions of nearby galaxies in which
there is no need for dark matter on these scales.  On large scales ($R/R_e\gg 1$)
there is no doubt that dark matter dominates, primarily based either on X-ray observations
of nearby early-type galaxies (e.g. Fabbiano~\cite{Fabbiano1989p87}, Lowenstein \& 
White~\cite{Lowenstein1999p50}) or weak lensing surveys
(e.g. Sheldon et al.~\cite{Sheldon2004p2544}).  
On intermediate scales ($R/R_e \sim 1$--$10$), there is considerable evidence that
early-type galaxies show roughly the same conspiracy as late-type galaxies between
the decline in the baryonic contribution and the rise of the dark matter 
contribution so as to have mass distributions that come close to producing
a flat rotation curve (e.g. Rusin et al.~\cite{Rusin2003p29}).  However, there are recent data from
the halo dynamics of nearby galaxies (Romanowsky et al.~\cite{Romanowsky2003p1396}),
time delay measurements in gravitational lenses (Kochanek~\cite{Kochanek2002p25}), 
and stellar dynamics in gravitational lenses (Treu \& Koopmans~\cite{Treu2004p739}) 
suggesting that this standard picture is either wrong and early-type galaxies 
have little dark matter in this region or that the density structure of early-type
galaxies is very heterogeneous.  The latter possibility is difficult to
reconcile with the tightness of either the familiar dynamical fundamental
plane (e.g. Djorgovski \& Davis~\cite{Djorgovski1987p59}) 
or the equally tight fundamental plane of gravitational 
lenses (Kochanek et al.~\cite{Kochanek2000p131}).

Here we are going to focus on the ability of gravitational lenses
to clarify these problems.  In interpreting lens data we must start
with the dreaded ``lens model.''  For whatever reasons, these two
words induce a mysterious level of anxiety in the populace at large,
when the reality is that lens models are no more (or less) bizarre
beasts than the ``dynamical models'' whose complexities, structures and
degeneracies seem to be accepted with reasonable equanimity.  For lens
models, however, there is a strange belief that somehow 
data~$+$~model yields random noise.  Much of
this arises because we tend to fit the models without providing a clear
explanation of how lens models work, which quantities are constrained 
and how any degeneracies arise.  One objective of this review is to clearly 
explain how strong lens data constrain mass distributions.  In \S\ref{sec:lens} we review
how the most familiar type of lens data, image positions, constrain the
mass distribution.  Then in \S3--6 we explore how the statistics of 
the lens sample, stellar dynamics of lenses, time delay measurements and
monitoring for microlensing variability can be used to avoid the limitations
of constraints based only on image positions.  

\section{How Lenses Constrain Mass Distributions} \label{sec:lens}

The fundamental frustration of gravitational lenses is that they always measure
something very accurately, but that something is usually a degenerate
combination of two interesting quantities.
In this section we review how lens galaxies constrain the mass distribution.
In our discussion we will focus only on the radial mass distribution (the
monopole).  In actual lens models, the angular structure is always important, 
but for the determination of the monopole it is simply a source of ``noise.''
The role of angular structure is discussed in detail in Kochanek~(\cite{Kochanek2004a}), 
but it has no particular effect on the discussion which follows.  All 
lens modeling, parametric or non-parametric, obeys the rules we are
about to describe.  Unfortunately, the gravitational lens community seems
to be more interested in obfuscating than explaining what it is that lens data
constrain (mostly to our own detriment).

\begin{figure}[t]
\begin{center}
\centerline{
      \includegraphics[width=2.5in]{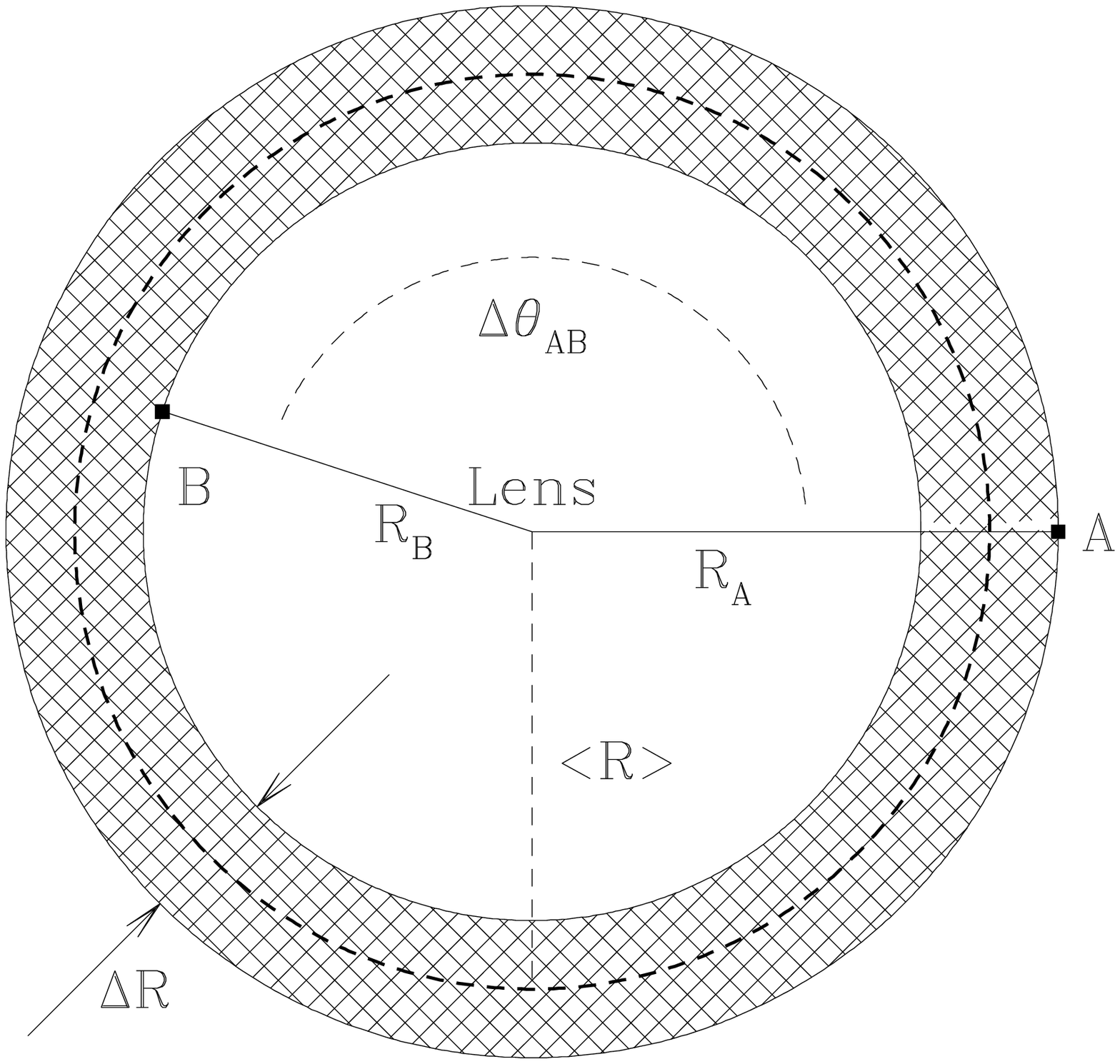}
      \includegraphics[width=2.5in]{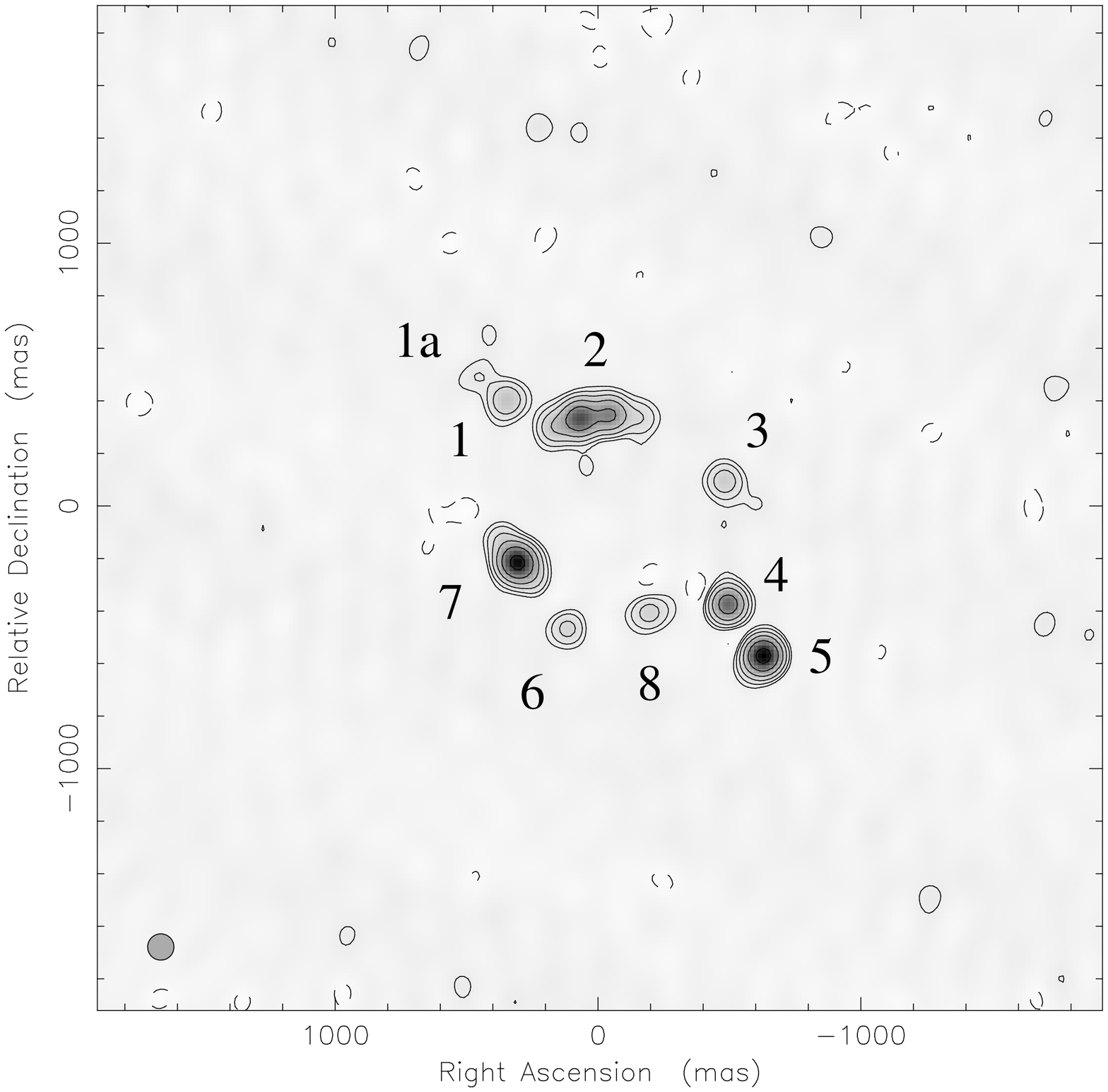}
      }
\end{center}
\caption{ (Left)
  A schematic diagram of a two-image lens.  The lens galaxy lies at the origin
  with two images A and B at radii $R_A$ and $R_B$ from the lens center.  The images
  define an annulus of average radius $\rbar=(R_A+R_B)/2$ and width $\dr=R_A-R_B$
  and they subtend an angle $\Delta\theta_{AB}$ relative to the lens center.  For
  a circular lens $\Delta\theta_{AB}=180^\circ$ by symmetry.
  }
\label{fig:geometry}
\caption{(Right) A Merlin map of B1933+503 showing the 10 observed images of
  the three component source (Marlow et al.~\cite{Marlow1999p15}).
  The flat radio spectrum core is lensed into images 1, 3, 4 and 6.
  One radio lobe is lensed into images 1a and 8, while the other is
  lensed into images 2, 7 and 5.  Image 2 is really two images
  merging on a fold.
  }
 \label{fig:b1933merlin}
\end{figure}

\begin{figure}[t]
\begin{center}
\centerline{\includegraphics[width=2.5in]{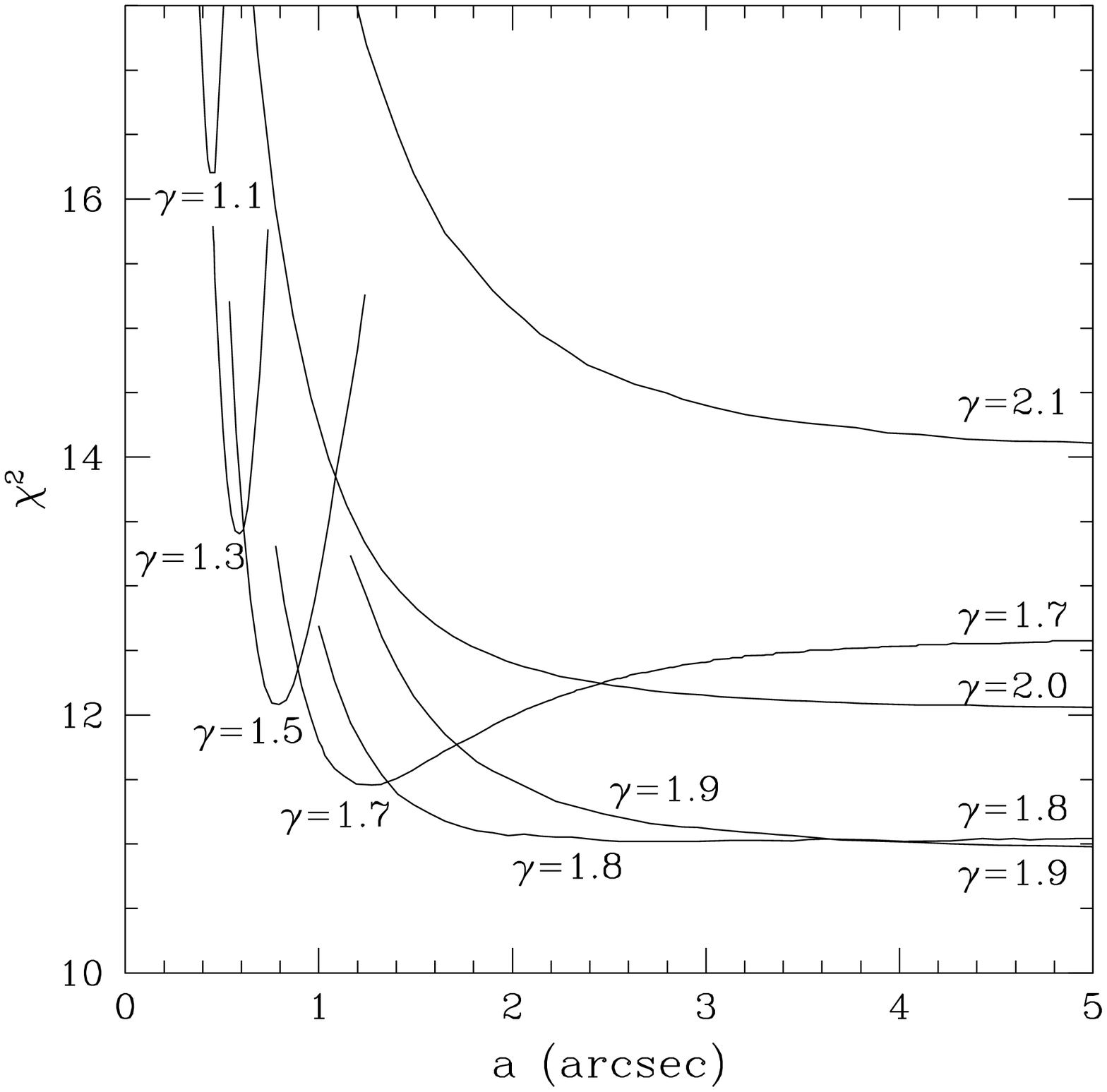}
            \includegraphics[width=2.5in]{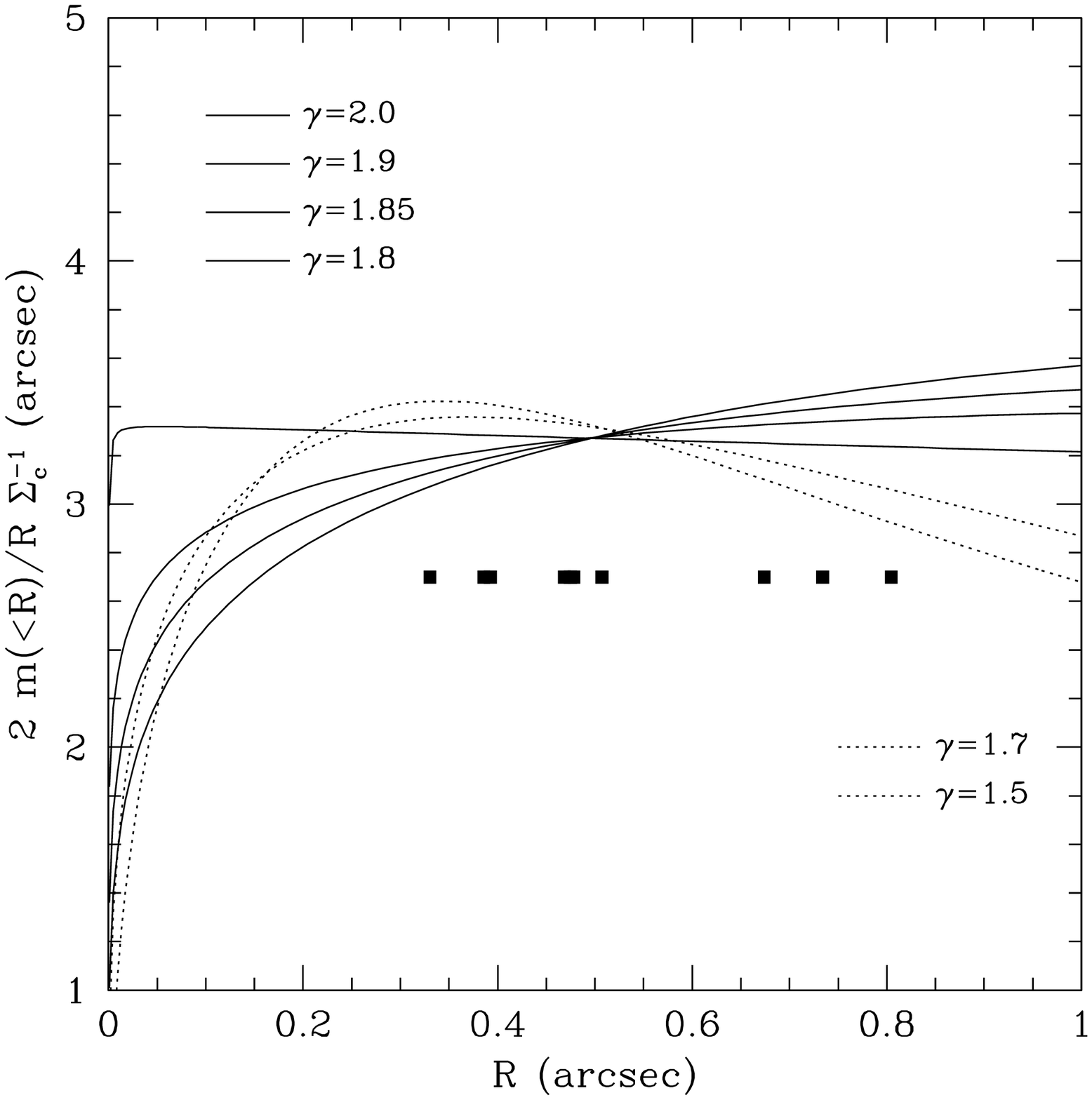}}
\end{center}
\caption{(RIGHT)
   Goodness of fit $\chi^2$ for cuspy models of B1933+503 as a function of the
   inner density exponent $\gamma$ ($\rho\propto r^{-\gamma}$) and the profile
   break radius $a$.  Models with cusps significantly shallower or steeper
   than isothermal are ruled out, and acceptable models near isothermal must
   have break radii outside the region with the lensed images.
   (LEFT)
   The monopole deflections of the B1933+503 models for the range of permitted
   cusp exponents $\gamma$.  The points show the radii of the lensed images,
   and the models only constrain the shape of the monopole in this region.
   The monopole deflection is closely related to the square of the rotation
   curve.  
   }
\label{fig:mod1933a}
\end{figure}

\begin{figure}[t]
\begin{center}
\centerline{\includegraphics[width=2.5in]{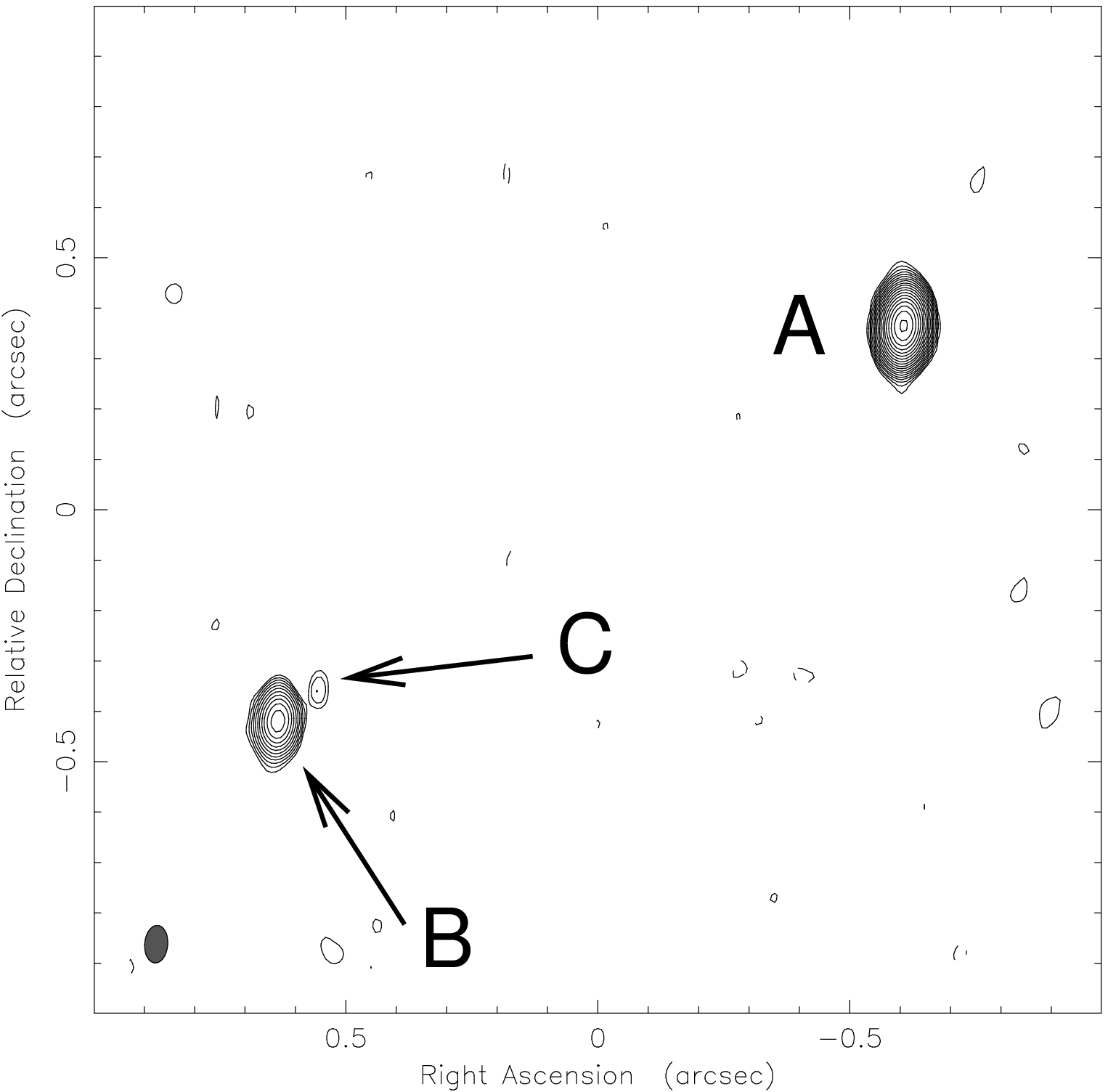}
            \includegraphics[width=2.5in]{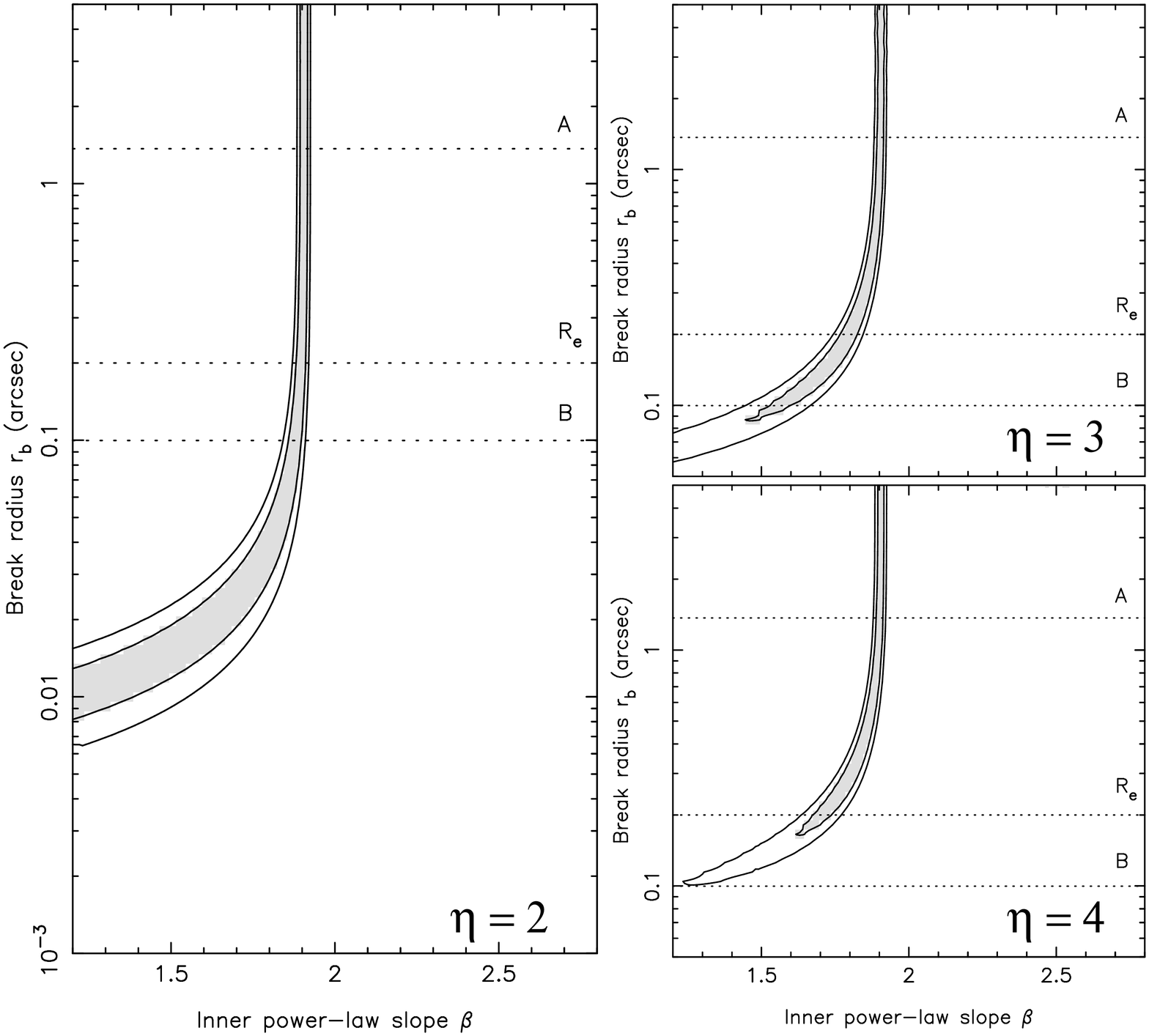}}
\end{center}
\caption{(RIGHT)
  PMN1632--0033 is the only known lens with a ``classical'' third
  image in the core of the lens galaxy.  The center of the lens galaxy
  is close to the faint C image.  Images A, B and C have identical radio
  spectra except for the longest wavelength flux of C, which can be
  explained by absorption in the core of the lens galaxy.  Time delay
  measurements would be required to make the case absolutely secure.
  A central black hole in the lens galaxy might produce an additional
  image with a flux about 10\% that of image C. (Winn et al.~\cite{Winn2004p613})
  }
\label{fig:pmnj1632}
\caption{
   (LEFT)
     Allowed parameters for cuspy models of PMNJ1632--0033 assuming that image
   C is a true third image.  Each panel shows the constraints on the inner
   density cusp $\beta$ ($\rho \propto r^{-\beta}$) and the break radius
   $r_b$ for three different asymptotic density slopes $\rho \propto r^{-\eta}$.
   A Hernquist model has $\beta=1$ and $\eta=4$, an NFW model has
   $\beta=1$ and $\eta=3$, and a pseudo-Jaffe model has $\beta=2$ and
   $\eta=4$.  Unless the break radius is place interior to the B image,
   it is restricted to be close to isothermal ($\beta=2$).
   }
\label{fig:mod1632}
\end{figure}

Fig.~\ref{fig:geometry} illustrates a simple gravitational lens with two images located 
at radii $R_A> R_B$ from the center of the lens galaxy.  Usually we assume
a mass distribution $M(<R)$ which produces a ray deflection 
$\alpha(R) =b \hat{\alpha}(R) \propto M(<R)/R$ consisting of an unknown normalization $b$
multiplied by a deflection profile determined by the mass distribution $\hat{\alpha}(R)$. 
Because both images must correspond to a common source, the lens 
equation sets the constraint that
\begin{equation}
            -R_B + b \hat{\alpha}(R_B) = R_A - b \hat{\alpha}(R_A)
   \label{eqn:constraint}
\end{equation}
which is easily solved to find a normalization of
\begin{equation}
            b = { R_A + R_B \over \hat{\alpha}(R_A) + \hat{\alpha}(R_B) }.
   \label{eqn:norm}
\end{equation}
Since the enclosed mass at any radius scales as $b\hat{\alpha}(R)$,
 you get an accurate estimate for the
enclosed mass independent of the profile.  In particular, if you use the mass inside the 
mean image radius, $\rbar=(R_A+R_B)/2$, then the enclosed mass is 
independent of the slope ($d\hat{\alpha}/dR$) of the deflection profile and has a 
fractional dependence on the profile of
\begin{equation}
    { \Delta M \over M } = {1 \over 4 } 
    \left[ { \rbar^2 \over \hat{\alpha} } { d^2\hat{\alpha} \over dR^2 } \right]
     { \dr^2 \over \rbar^2 } 
\end{equation}
that is second order in the asymmetry $\dr/\rbar$ of the image positions.  As the
lens becomes symmetric, $R_A \rightarrow R_B$, the images are approaching
the Einstein radius and the mass uncertainties diminish rapidly because 
for any circular lens the mean surface density of the material interior
to the Einstein radius $R_E$ is exactly $M=\pi R_E^2 \Sigma_c$ where
$\Sigma_c$ is the critical surface density for lensing.  

The mass normalization (Eqn.~\ref{eqn:norm}) is not very illuminating if
our objective is to understand the property of the density distribution
constrained by the lensed images.  To understand the constraint 
we start by introducing the mean surface density in an annulus,
\begin{equation}
           \langle \kappa_{12} \rangle 
                = { 2 \int_{R_1}^{R_2} R \kappa(R) dR \over R_2^2-R_1^2 },
\end{equation} 
where the mean surface density is related to the mass by 
$M_{12} = \pi \langle \kappa_{12} \rangle \Sigma_c (R_2^2-R_1^2)$.
The ray deflections at the two images are simply
\begin{equation}
  b \hat{\alpha}(R_B) = { 1 \over R_B } \langle \kappa_{0B} \rangle R_B^2
\qquad
  b \hat{\alpha}(R_A) = { 1 \over R_A } \left[ \langle \kappa_{0B} \rangle R_B^2 + 
      \langle \kappa_{AB} \rangle (R_A^2-R_B^2)\right].
\end{equation}
Substituting these into the constraint equation (Eqn.~\ref{eqn:constraint}) we find that
\begin{equation}
        { 1 - \langle \kappa_{0B} \rangle \over 1-\langle\kappa_{AB} \rangle } =
         {R_B - R_A \over R_B } < 0.
        \label{eqn:onezone}
\end{equation}
The positions of the two images measure a combination of the mean surface
density interior to the inner image and the mean surface density between the
two images to the accuracy of the astrometric measurements.  This will still
hold if we add angular structure to the lens model if we think of the 
angular deflections as an added source of noise in the constraint.  
The surface densities are constrained to the range
\begin{equation}
  0 \leq \langle \kappa_{AB} \rangle \leq 1 \leq \langle \kappa_{0B} \rangle 
    \leq \kappa_{max}=R_A/R_B
\end{equation}
with a linear trade off 
\begin{equation}
     \langle \kappa_{0B} \rangle = \kappa_{max} \left(1-\langle\kappa_{AB} \rangle\right)
    + \langle\kappa_{AB} \rangle
\end{equation}
between the surface densities in which the mass interior to B decreases as we 
add mass to the annulus between A and B.

The particular combination of surface densities on the
left side of Eqn.~\ref{eqn:onezone} arises because of the mass sheet 
degeneracy (Falco et al.~\cite{Falco1985p1}) -- given a surface density $\kappa(r)$, the image positions
and flux ratios are unchanged if we transform the surface density to
by $\kappa(r) \rightarrow (1-\kappa_0) \kappa(r) + \kappa_0$.  It is called
the mass sheet degeneracy because it corresponds to adding a sheet of 
surface density $\kappa_0$ while reducing the mass scale of the original 
profile by the factor $1-\kappa_0$.  Since the
right side of Eqn.~\ref{eqn:onezone} is a measured quantity, the left side   
must be invariant under this transformation.  

If the lens has additional images in the interior of the annulus, then we
can measure more properties of the surface density.  The simplest example
is to suppose that in addition to the images at $R_A$ and $R_B$ we also
see an Einstein ring image in the annulus at $R_B < R_E < R_A$.  As 
discussed earlier, this is the unique point where the enclosed mass is
determined uniquely, with $\langle \kappa_{0E} \rangle = 1$ in a 
spherical lens.  The added constraint allows us to determine the  ratio 
\begin{equation}
        { 1 - \langle \kappa_{BE} \rangle \over 1-\langle\kappa_{EA} \rangle } =
        {R_B \over R_A }  {R_A^2 - R_E^2 \over R_E^2 - R_B^2} > 0.
        \label{eqn:twozone}
\end{equation}
which relates the surface density of the annulus inside the Einstein ring
to that outside the Einstein ring.  Once again, the constrained quantity
combines two mean surface densities in a form which is invariant under
the transformation associated with the mass sheet degeneracy.

This brings us to the Gauss' Laws for (spherical) lens models given a lens
with images bounded by the annulus $R_B < R < R_A$:
\begin{itemize}
\item Only the total mass interior to $R_B$ and not its distribution is relevant,
\item The distribution of mass in the annulus $R_B < R < R_A$ is relevant only
  if there are additional lensed images in the annulus, and
\item The amount and distribution of mass exterior to $R_A$ is irrelevant.
\end{itemize}
Parametric models must have enough degrees of freedom to vary these quantities,
but the constraints on the profile shape apply only in the annulus and only
to the extent there are additional constraints.  Any extra degrees of freedom in a
model beyond the mean annular densities are irrelevant.

If we have a lens like B1933+503 (Sykes et al.~\cite{Sykes1998p310}) with 10 lensed 
images produced by three source components (see Fig.~\ref{fig:b1933merlin}), then the 
density structure in the annulus can be tightly constrained.  Fig.~\ref{fig:mod1933a}
shows the constraints obtained by Mu\~noz et al.~(\cite{Munoz2001p657}, also see
Cohn et al.~\cite{Cohn2001p1216}) on density distributions of the form 
$\rho \propto r^{-n} (r^2 +a^2)^{1+n/2}$.  The acceptable models
have parameters ($n \simeq 2$ and $a >> R_E$) that lead to nearly flat 
rotation curves over the annulus bounding the lensed images.  Other recent
examples are analyses of B0218+357 by Wucknitz et al.~(\cite{Wucknitz2004p14})    
and of 0047--2808 by Dye \& Warren~(\cite{Dye2004p1}).

\begin{figure}[t]
\begin{center}
\centerline{\includegraphics[width=5.0in]{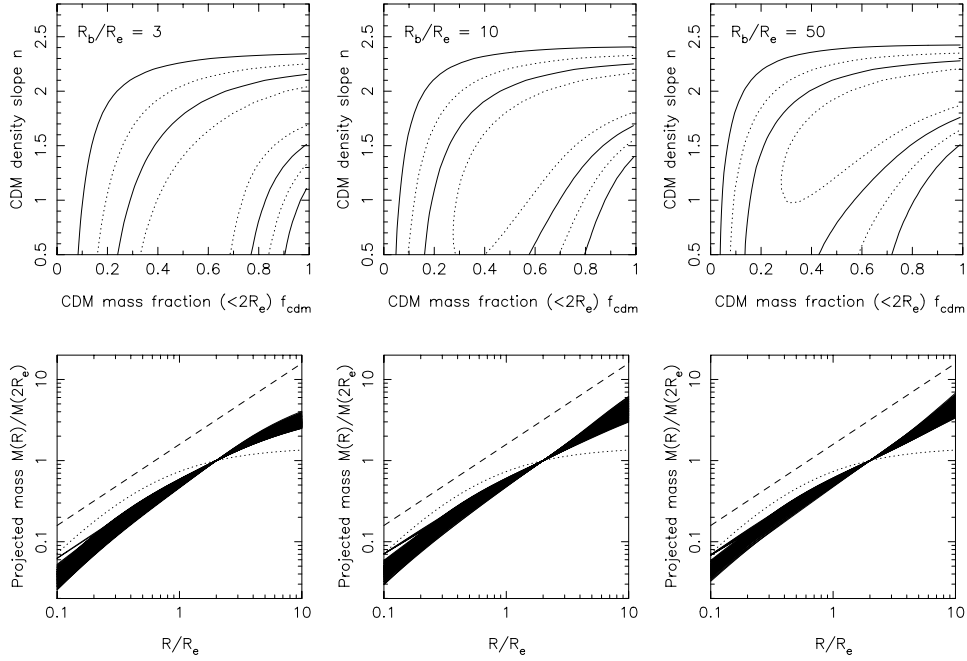}}
\end{center}
\caption{
   The structure of lens galaxies in self-similar models.  The top row
   shows the permitted region for the slope of the inner dark matter cusp
   ($\rho \propto r^{-n}$) and the projected fraction of the mass $f_{CDM}$ inside $2R_e$
   composed of dark matter.  The results are shown for three ratios $R_b/R_e$
   between the break radius $R_b$ of the dark matter profile and the effective
   radius $R_e$ of the luminous galaxy.  The solid (dashed) contours show the 68\% and 95\%
   confidence levels for two (one) parameter.  Note that the estimates of
   $n$ and $f_{CDM}$ depend little on the location of the break radius relative
   to the effective radius.  The bottom row shows all the mass profiles lying
   within the (two parameter) 68\% confidence region normalized to a fixed projected
   mass inside $2R_e$.  For comparison we show the mass enclosed by a de Vaucouleurs
   model (dotted line) and an SIS (offset dashed line).  While the allowed models
   exhibit a wide range of dark matter abundances, slopes and break radii, they
   all have roughly isothermal total mass profiles over the radial range spanned
   by the lensed images.
   }
 \label{fig:selfsim}
\end{figure}

Another example of an additional constraint is if we can detect a central or
odd image at radius $R_C < R_B$.  While there are few examples of central
images as yet, their detection will become routine as the sensitivity
of radio observations increase.  Carrying out the same analysis in terms
of mean surface densities that lead to Eqn.~\ref{eqn:onezone}, we find the
constraint 
\begin{equation}
    { 1 - \langle \kappa_{0C} \rangle \over 1 -\langle \kappa_{BC} \rangle}
     = 1 + {R_B \over R_C }
\end{equation}
between the mean density $\langle \kappa_{0C} \rangle$ interior to image 
C and the mean density $\langle \kappa_{BC} \rangle$ between images 
B and C.  If we have a very asymmetric triple in which 
$\langle \kappa_{0C} \rangle \gg 1$ and $\langle \kappa_{BC} \rangle \gg 1$,
then the central density is larger by the factor $1+R_B/R_C$.
The constraint in Eqn.~\ref{eqn:onezone} also holds, and
it is easy to derive the constraints on other combinations of the mean
surface densities to find that they all have the standard form with
ratios of $1-\langle\kappa\rangle$ equal to an astrometric constraint.
The cleanest present example of a lens with a third image is
the radio lens PMNJ1632--0033 (Winn et al.~\cite{Winn2002p10},
Winn, Rusin \& Kochanek~\cite{Winn2004p613}) shown in Fig.~\ref{fig:pmnj1632}.
Like B1933+503, the mass distribution is constrained 
the mass profile is required to be close to isothermal with 
$1.89 < n < 1.93$ (see Fig.~\ref{fig:mod1632}).  Routine detection
of these central images also offers a new probe for black holes
at the centers of the lens galaxy (Mao et al.~\cite{Mao2001p301}) --
in the case of PMNJ1632--0033 most models with a black hole as 
massive as expected from local scalings will produce a fourth
image (D) close to image C but about 10 times fainter.

Unfortunately, lenses with additional constraints are relatively rare, so we must find 
alternate means of breaking the degeneracy.  Einstein ring images of the host galaxy
can be found relatively easily, but they are better suited for constraining the
angular structure of the lens than the radial structure.
For lenses without additional constraints 
this can be done statistically provided the lenses have relatively homogeneous mass
distributions.  For individual lenses without additional constraints we must either 
measure the mass  $\langle \kappa_{OB}\rangle$ in the interior or measure the surface 
density of the annulus $\langle \kappa_{AB}\rangle$. 
The lens constraint Eqn.~\ref{eqn:onezone} then supplies the other surface density
and we have an absolute measurement of the profile for $R < R_A$.  A very different
approach is to break the degeneracy by using microlensing variability to determine
the ratio $\kappa_*/\kappa$ between the surface density in stars and the total 
surface density in the annulus.

\section{Statistical Constraints on the Mass Distribution}

Analyzing ensembles of lenses to estimate the mean mass distribution has two
advantages over analyzing individual lenses.  First, it provides constraints
over a broad range of radii (relative to $R_e$), and second, it breaks the
mass sheet degeneracy.  The disadvantage of the method is that we must assume
the density distributions of lenses are self-similar or at least heterogeneous
in a manner that can be included as part of the model.  
The mass sheet degeneracy is broken because all lenses must
agree on the same (scaled) physical density distribution but have source and
lens redshifts corresponding to different critical densities.  Thus, adding
a constant physical density sheet corresponds to adding a different critical
density sheet to each lens which cannot be reconciled with the constraints.
This approach has been used by Rusin et al.~(\cite{Rusin2003p29},
\cite{Rusin2004p1}) to estimate both the average mass distribution and the
evolution of stellar populations.  

Rusin et al.~(\cite{Rusin2004p1}) model the lens galaxy as the sum of a 
Hernquist model normalized by the observed effective radius of the lens
galaxy embedded in a dark matter halo of the form 
\begin{equation}
   \rho = { \rho_c \over (r/r_b)^n \left[1+(r/r_b)^2\right]^{(3-n)/2} }
\end{equation}
were $r_b$ is the break radius and $n$ is the logarithmic slope of the
central density cusp.  Here we are interested in the dark matter slope $n$
and the fraction of the projected mass inside $2R_e$ which is dark matter, 
although the full calculation includes variables for the scaling of galaxy
mass-to-light ratios with redshift and mass.  Fig.~\ref{fig:selfsim} shows the 
resulting constraints on the CDM mass fraction projected inside $2R_e$ and the logarithmic
slope of the central density cusp of the dark
matter.  While there is a degeneracy in the exact trade-off between the dark
matter fraction and the slope, the total mass distribution is well-constrained
and close to isothermal (flat rotation curve).  The residuals about this
mean mass profile can be entirely explained by a realistic spread in the
mean stellar ages of the lens galaxies. 

\begin{figure}[t]
\centerline{ \includegraphics[width=1.5in]{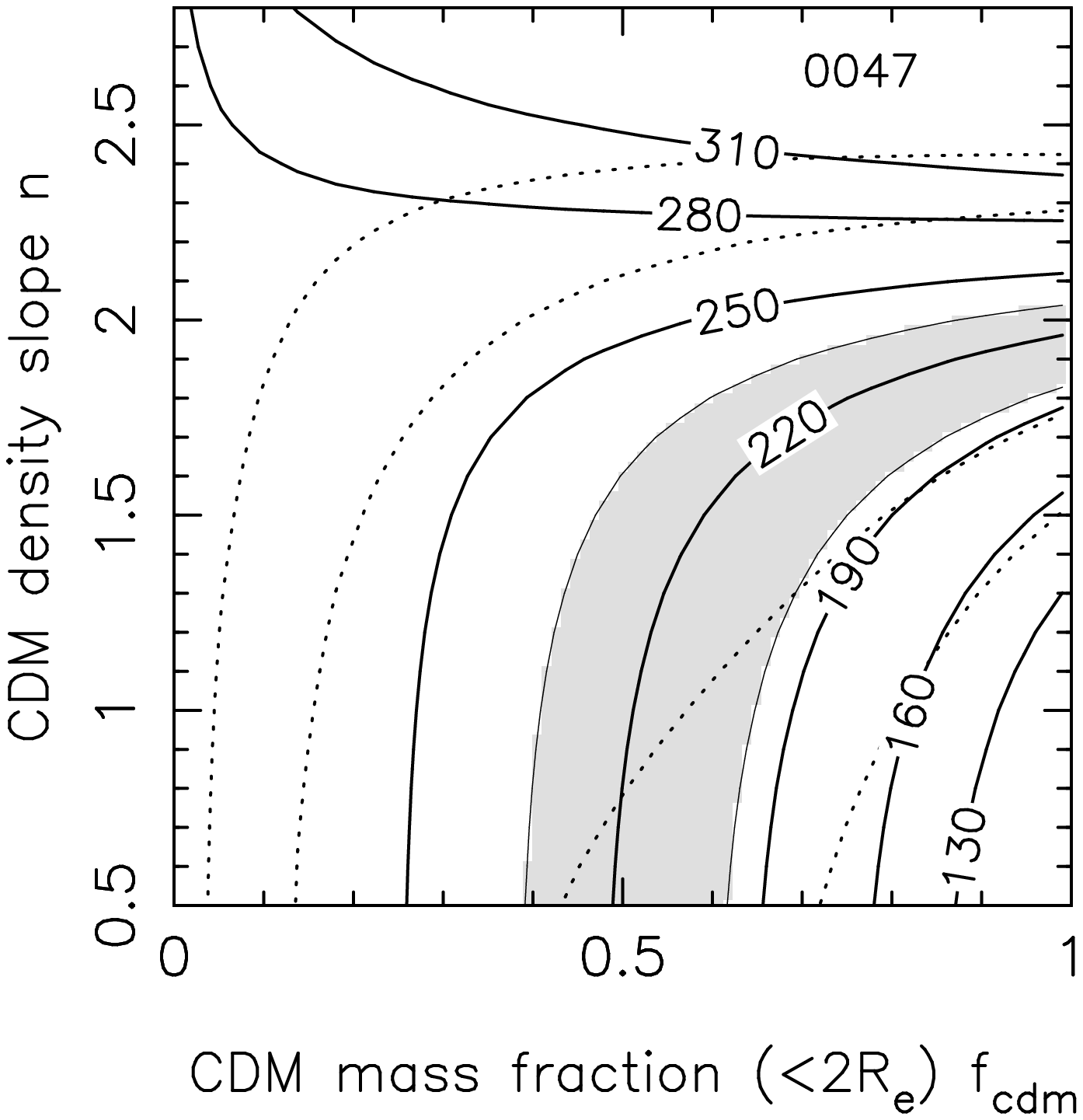}
             \includegraphics[width=1.5in]{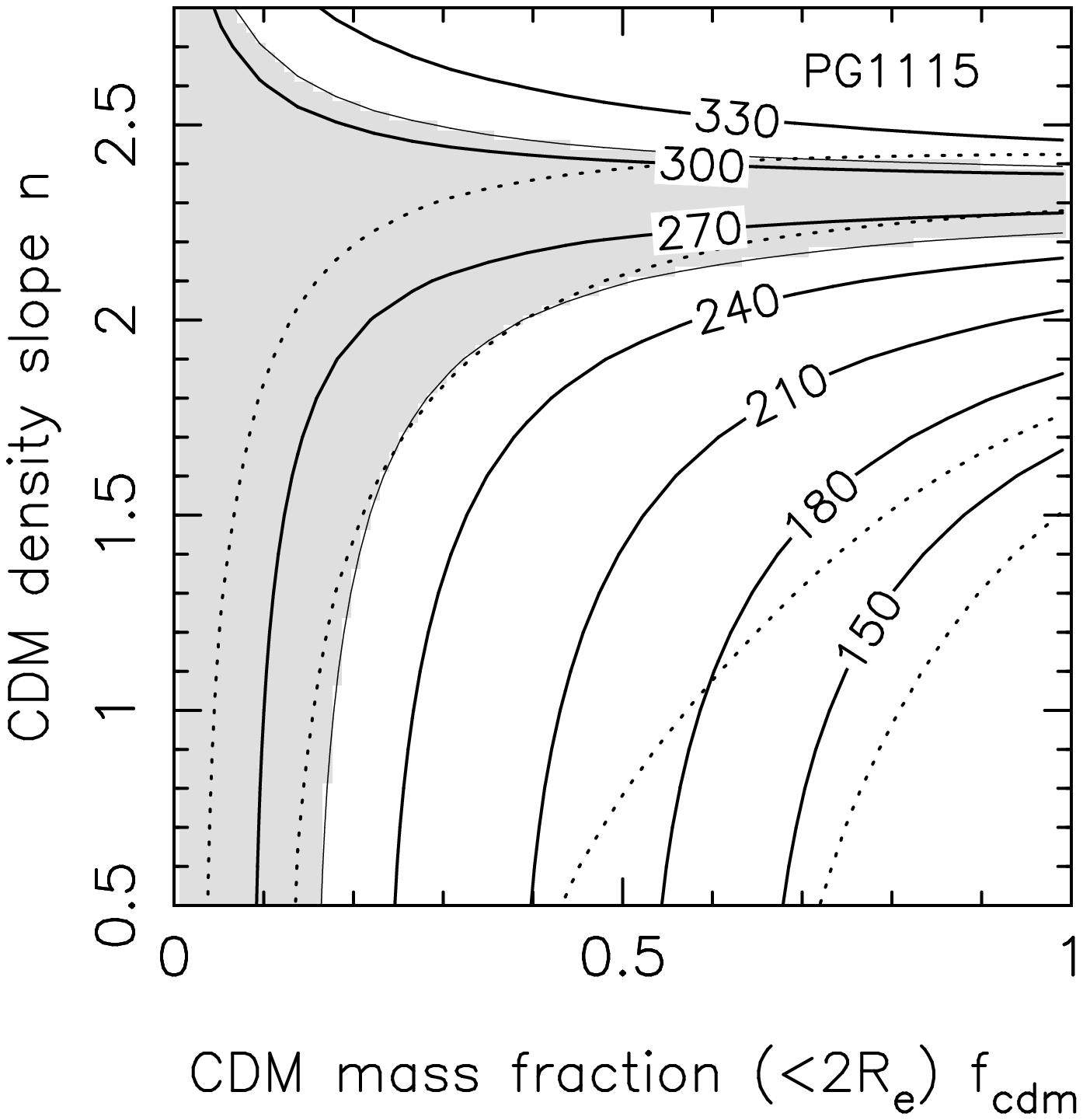}
             \includegraphics[width=1.5in]{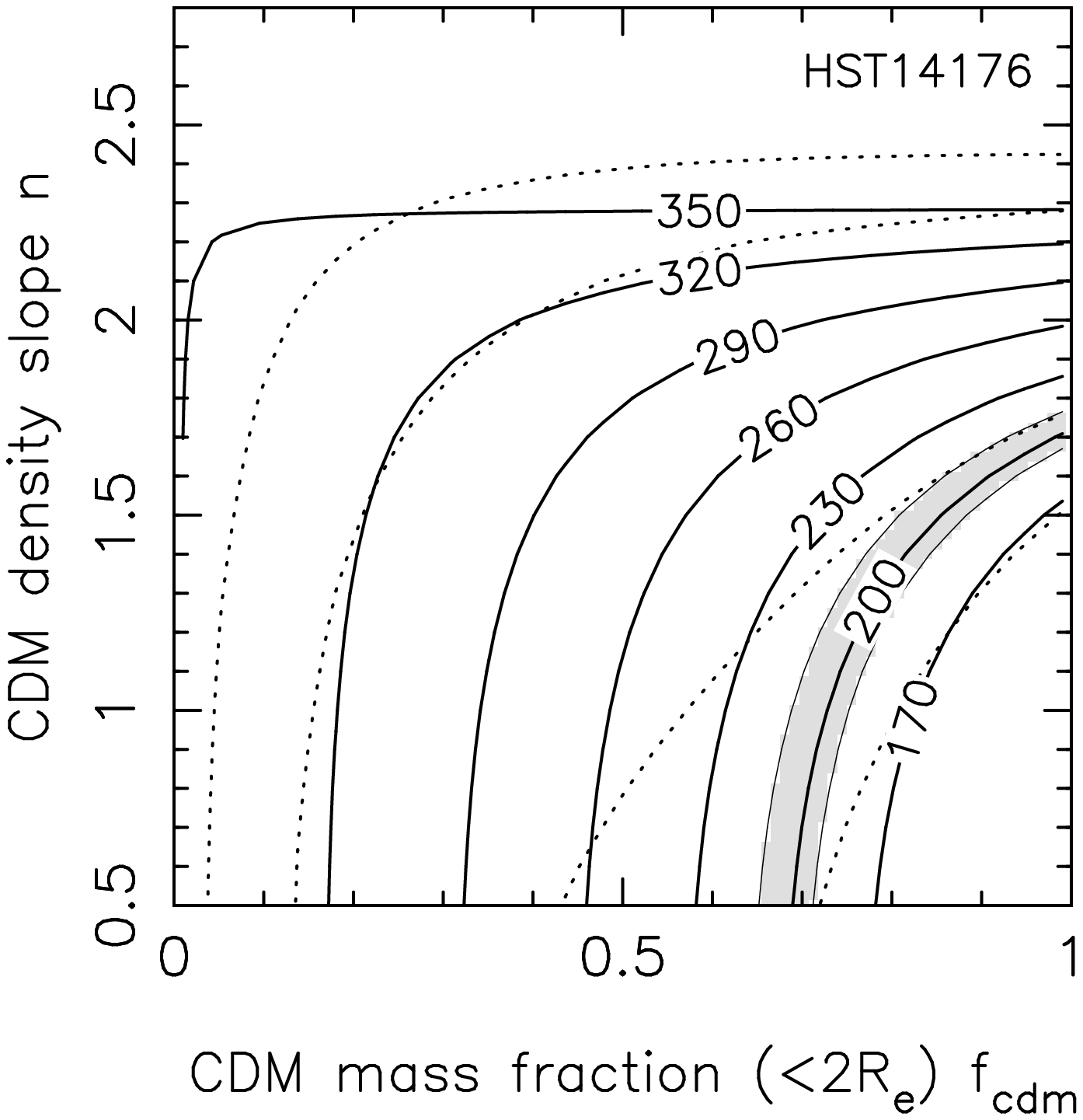} }
\centerline{ \includegraphics[width=1.5in]{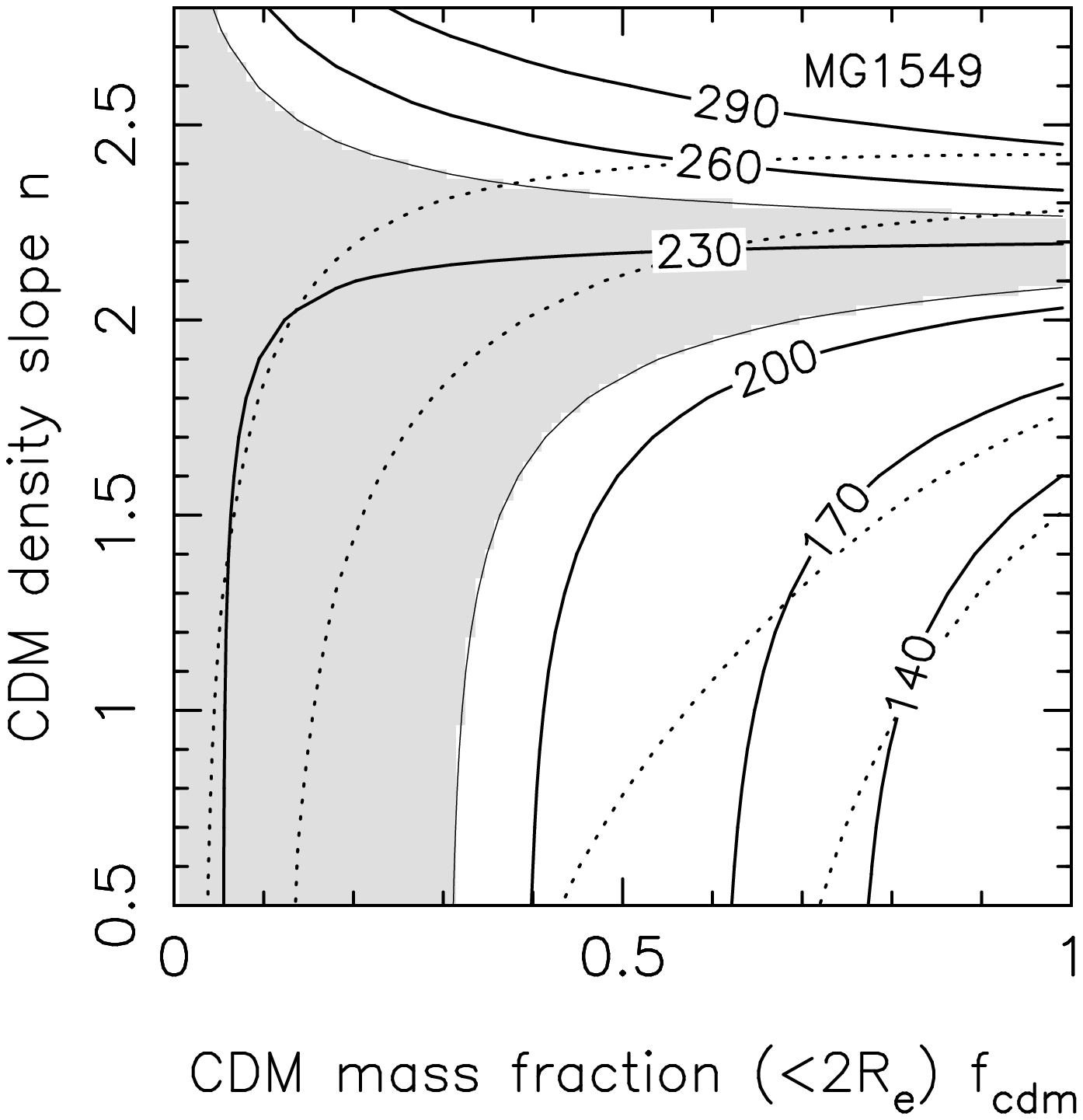}
             \includegraphics[width=1.5in]{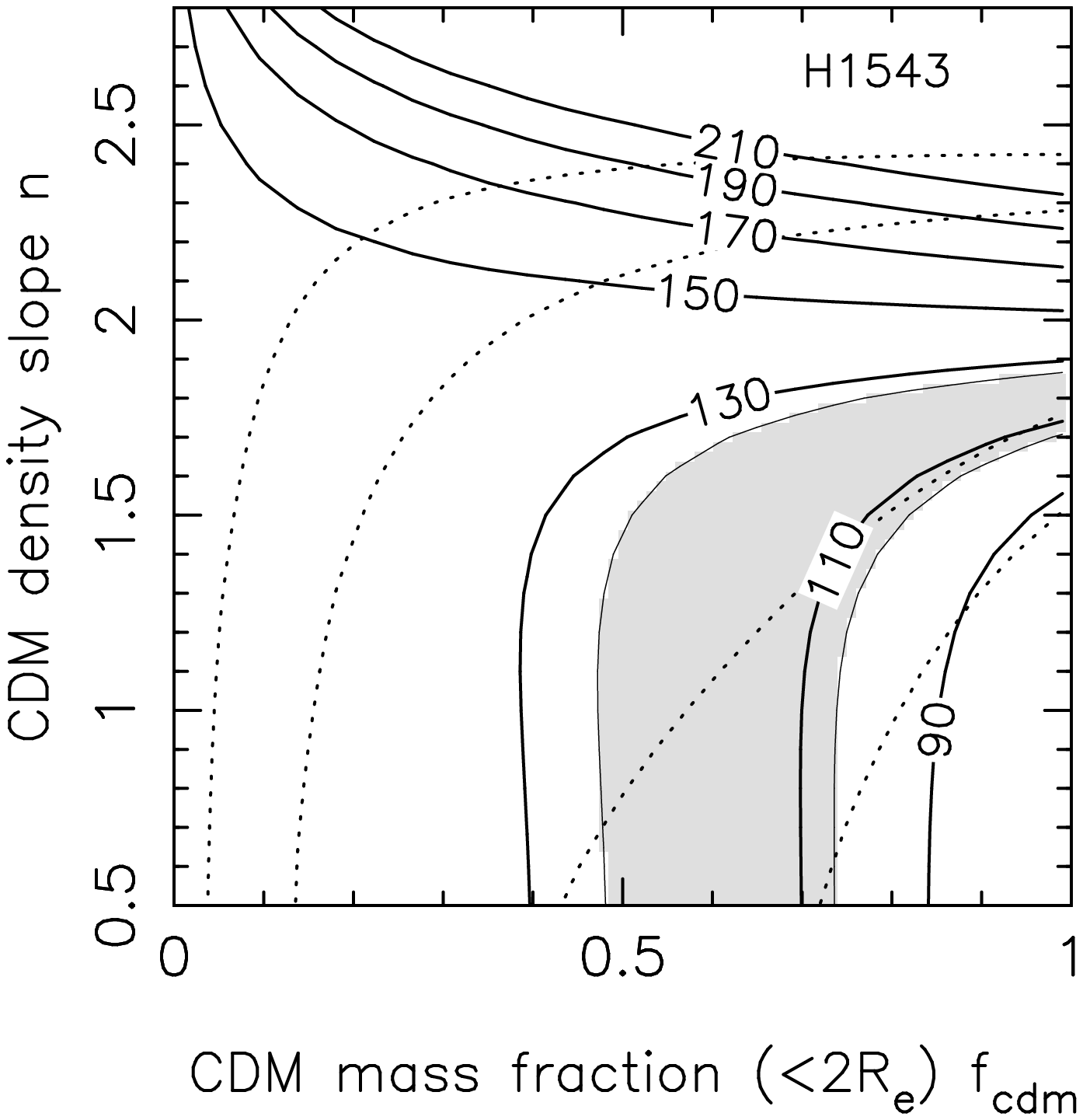}
             \includegraphics[width=1.5in]{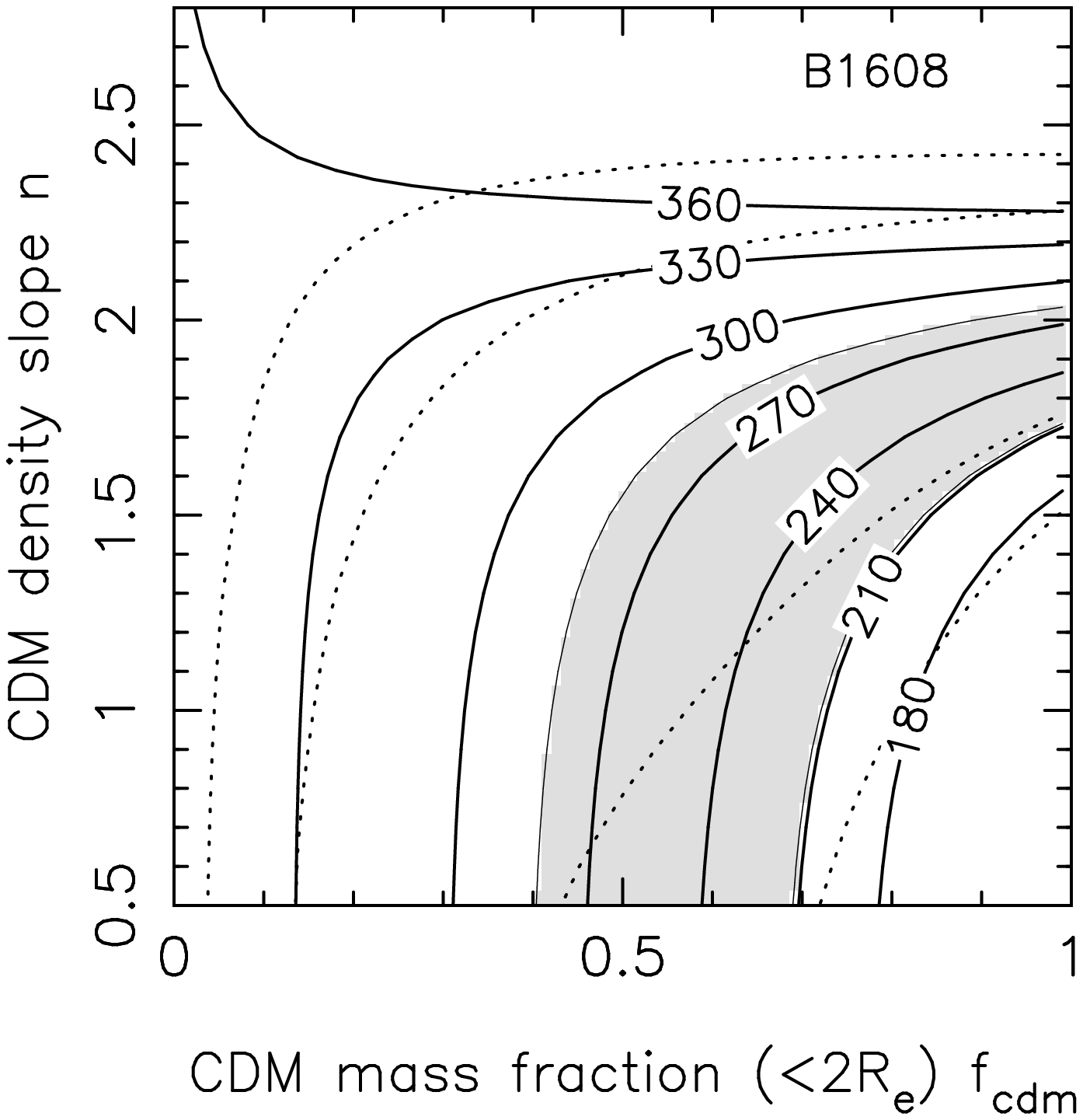} }
\centerline{ \includegraphics[width=1.5in]{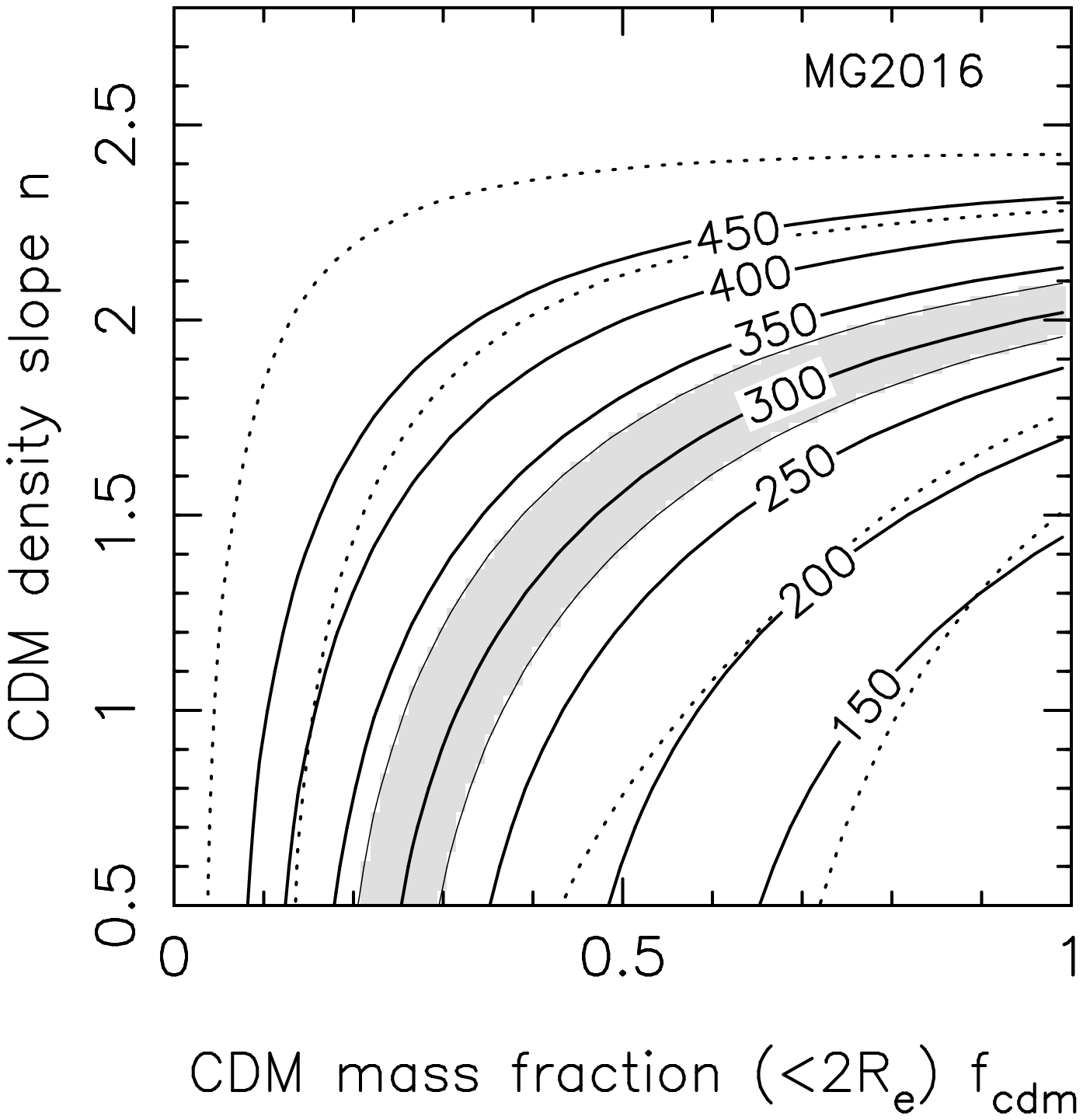}
             \includegraphics[width=1.5in]{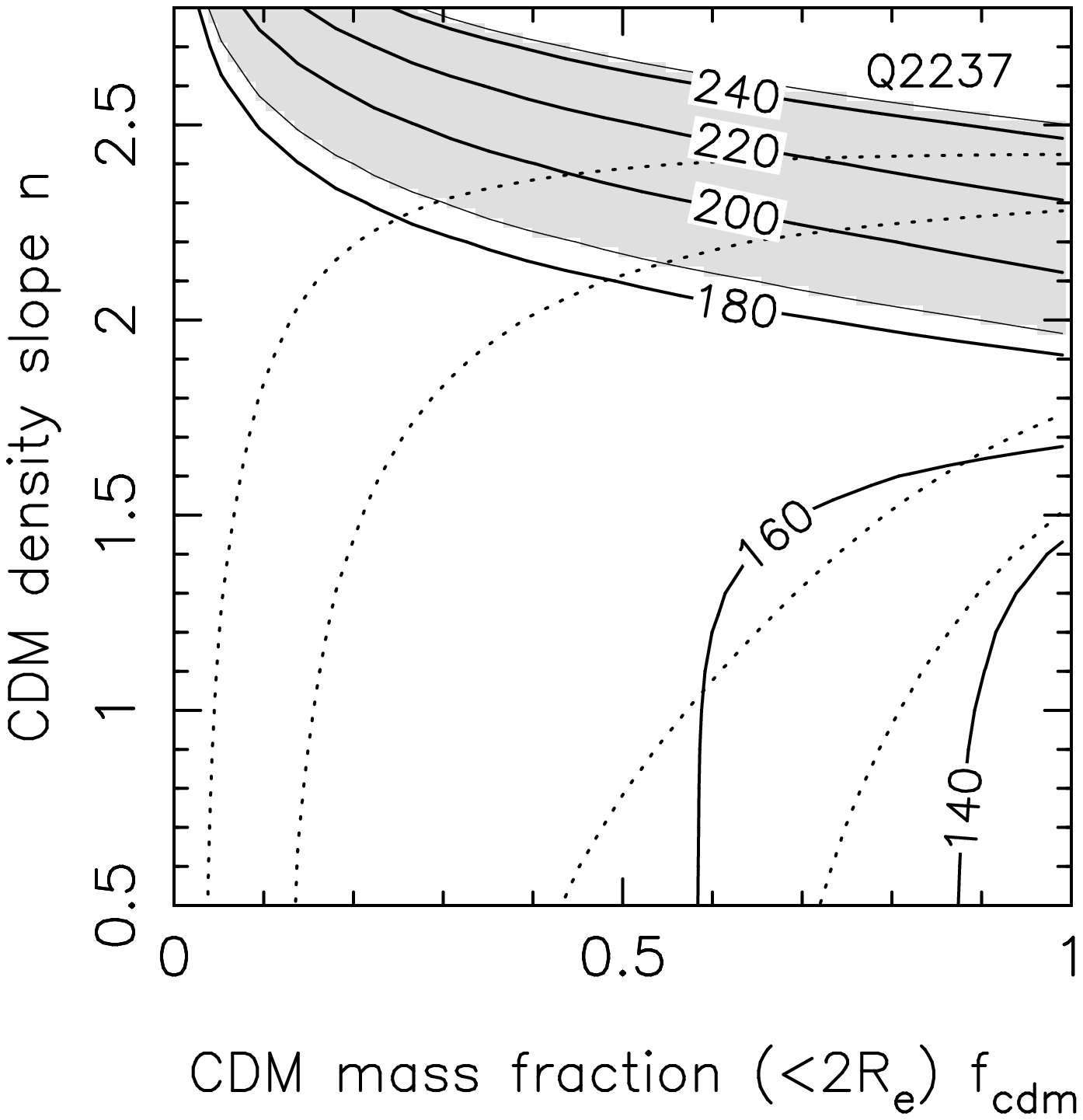}
             \includegraphics[width=1.5in]{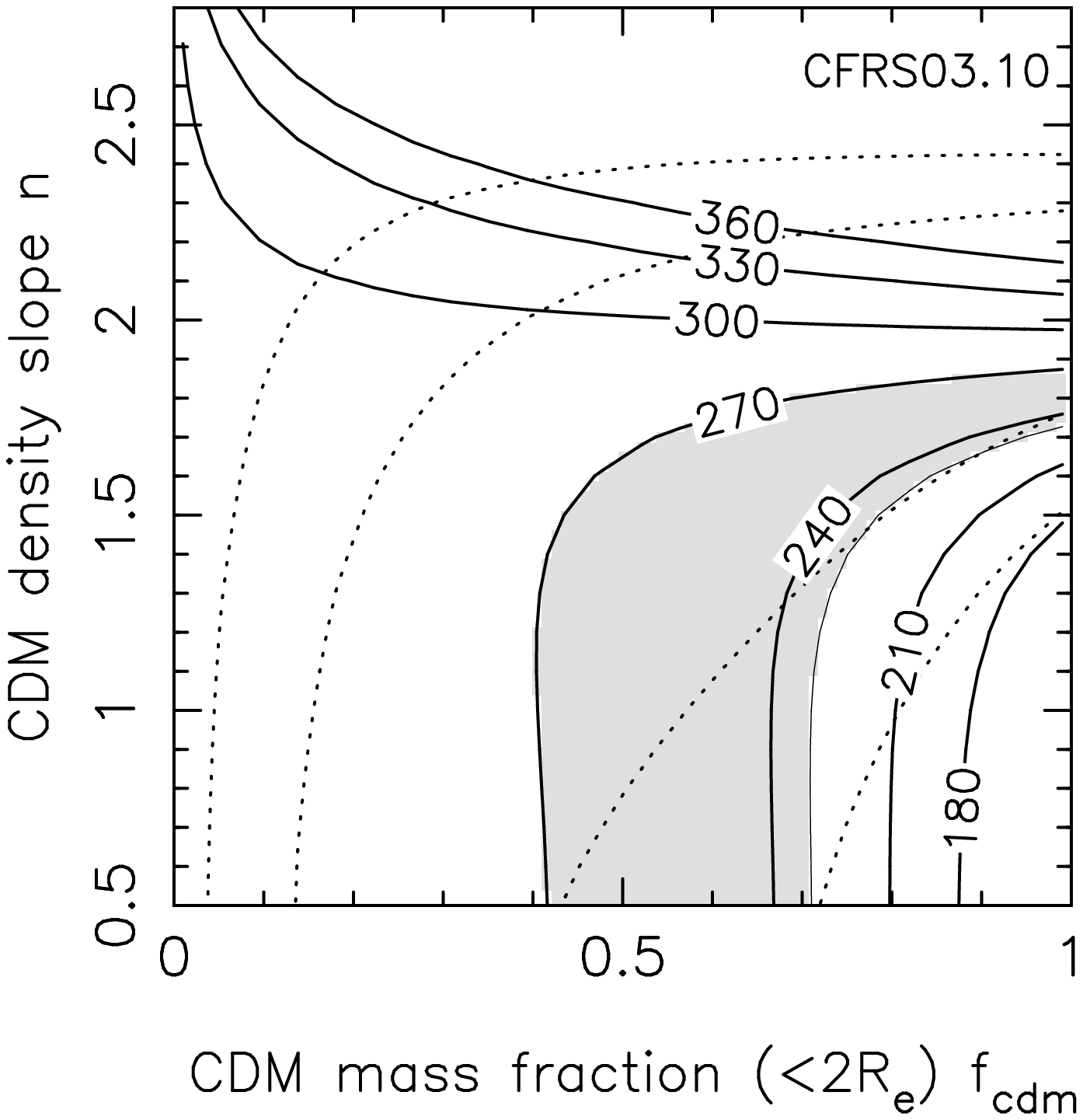} }
\caption{  Constraints from lens velocity dispersion measurements on the
   self-similar mass distributions of Fig.~\ref{fig:selfsim}. 
    The dotted contours show the 68\% and 95\%
   confidence limits from the self-similar models for $R_b/R_e=50$.
   The shaded regions show the models allowed (68\% confidence)
   by the formal velocity dispersion measurement errors, and the heavy solid
   lines show contours of the velocity dispersion in km/s.  We used
   the low Gebhardt et al.~(\cite{Gebhardt2003p1}) velocity dispersion
   for HST14176+5226 because it has the smallest formal error.  These
   models assumed isotropic orbits, thereby underestimating the full
   uncertainties in the stellar dynamical models.
   }
\label{fig:vdisp}
\end{figure}

\section{The Stellar Dynamics of Lens Galaxies \label{sec:vdisp}}     

If we can use another technique to measure either $\langle\kappa_{AB}\rangle$
or $\langle\kappa_{0B}\rangle$, then we can immediately use the basic 
lens constraint (Eqn.~\ref{eqn:onezone}) to determine the other mean
surface density and obtain a tight constraint on the density distribution. 
For example, measuring the stellar velocity 
dispersion of the lens galaxy will constrain the mass well inside the
Einstein radius where it is dominated by stars, roughly corresponding
to measuring $\langle\kappa_{0B}\rangle$. 

Another way to think about a dynamical measurement is to imagine that
the galaxy has a mass profile $M(R) = M_E (R/R_E)^{3-n}$ where
we know the mass $M_E$ inside the Einstein radius $R_E$ accurately
from the astrometry of the lensed images but cannot constrain the
density exponent $n$.  A velocity dispersion measurement 
corresponds to measuring a virial mass $M_v = c_v \sigma^2 R_v/G$
on the scale $R_v$ of the aperture with a dimensionless constant
$c_v$ compensating for projection effects, asphericity and 
orbital anisotropies.  Solving for the exponent we find that
\begin{equation}
    n = 3 - \left[ \ln(M_v/M_E) \right] \left[ \ln(R_v/R_E) \right]^{-1}
    \label{eqn:vscale}
\end{equation}
and we should be able to determine $n$ to an accuracy of
$\delta n \sim (1-2)(\delta\sigma/\sigma)$ since the ratio of
the radii is $|\ln(R_v/R_E)|\simeq 1-2$.

This has been done for 
10 systems\footnote{The systems are: 0047--2808 Koopmans \& Treu~\cite{Koopmans2003p606};
CFRS03.1077 Treu \& Koopmans~\cite{Treu2004p739};
Q0957+561 Falco et al.~\cite{Falco1997p70}, Tonry \& Franx~\cite{Tonry1999p512};
PG1115+080~Tonry~\cite{Tonry1998p1});
HST14176+5226 Ohyama et al.~\cite{Ohyama2002p2903}, Gebhardt et al.~\cite{Gebhardt2003p1},
   Treu \& Koopmans~\cite{Treu2004p739};
HST15433+5352  Treu \& Koopmans~\cite{Treu2004p739};
MG1549+3047 Leh\'ar et al.~\cite{Lehar1996p1812};
B1608+656 Koopmans et al.~\cite{Koopmans2003p70};
MG2016+112 Koopmans \& Treu~\cite{Koopmans2002p5}; and
Q2237+0305 Foltz et al.~\cite{Foltz1992p43}.}
 as illustrated in Fig.~\ref{fig:vdisp} using the same 
parameter space as used for the self-similar models in Fig.~\ref{fig:selfsim}.  With
the exception of Q2237+0305, where we see the images in the bulge of a nearby spiral
galaxy, the parameter space and degeneracies of the individual velocity dispersion
measurements are very similar to the results from the statistical analysis or the
models of individual lenses with extra constraints.  There 
are a few lenses which may require more centrally concentrated mass distributions
and a falling rotation curve (e.g. PG1115+080), and a few lenses which may require   
a less centrally concentrated mass distribution and a rising rotation curve
(e.g. HST14176+5226).  Treu \& Koopmans~(\cite{Treu2004p739}) summarize 
their individual estimates of the mean logarithmic slope of the density finding
results scattered about slightly rising rotation curve $\langle n\rangle \simeq 1.9$  
with an appreciable rms scatter of $0.30$.  The error estimates for the
effective slope $n$ seem to be 2--4 times smaller than expected from
our simple scaling argument in Eqn.~\ref{eqn:vscale}.
The finding that the typical lens has
a rising rotation curve is surprising given the expected halo 
structure shown in Fig.~\ref{fig:halo}, where we would expect to see a slowly
falling rotation curve for $R \gtorder R_e/2$ independent of the baryon fraction.

\def\vlos{\langle v_{los}^2 \rangle^{1/2}}
From Fig.~\ref{fig:vdisp} we might naively conclude that the stellar dynamical approach
is superior to the statistical approach.  However, in producing this comparison, 
Rusin et al.~(\cite{Rusin2004p1}) assumed a simple isotropic dynamical model and
the formal uncertainties, which leads to two problems.  The first problem is that dynamical
measurement uncertainties tend to be underestimated -- for example, the three velocity
dispersion measurements for HST14176+5226 can be mutually consistent only if their
error bars are expanded by 30\%.  The second problem is that several systematic 
uncertainties have yet to be included in the comparison.  These arise because 
the distribution of stellar velocities is not Gaussian and because galaxies are
not spheres, leading to systematic problems when we use the spherical Jeans equations
to analyze the data.  The systematics are different from local stellar dynamical
estimates because the profile slope (e.g. Eqn.~\ref{eqn:vscale}) compares a 
dynamical mass to a lensing mass rather than comparing two dynamical masses.
This makes estimates for the mass distribution using the stellar dynamics of 
lenses sensitive to simple calibration errors $\vlos = f \sigma_m$ with 
$f \neq 1$ between the measurement $\sigma_m$ and the rms velocity $\vlos$ 
in the Jeans equations.  Such errors cancel (to lowest order) in a purely
stellar dynamical estimate of the mass distribution.

Two simple examples are the effects of non-Gaussian line-of-sight velocity distributions (LOSVDs)
and deviations from spherical symmetry.  While the Jeans equations depend
on $\vlos$, the measurement is usually the dispersion of the best-fit Gaussian,
and these two quantities differ by $\vlos\simeq \sigma(1+\sqrt{6} h_4)$
where $|h_4|\ltorder 0.03$ is the dimensionless coefficient of an expansion of
the LOSVD in Gauss-Hermite polynomials (e.g. van der Marel \& Franx~\cite{van_der_Marel1993p525}).
The resulting corrections, $|f-1|\simeq \sqrt{6}h_4 \simeq 7\%$, can be comparable
to the formal measurement errors.  For example, Romanowsky \& Kochanek~(\cite{Romanowsky1999p18})
were able to find models of PG1115+080 consistent with a flat rotation curve where
Treu \& Koopmans~(\cite{Treu2002p6}), who used only the Jeans equations, could not.
Arguably the distribution functions required to produce the agreement were unusual,
but unusual is a very different statement from impossible.
Similarly, galaxies are not spheres.  From the
virial theorem it is easy to show that an oblate ellipsoid of axis ratio
$q$ has a ratio between the major and minor axis velocity dispersions that
corresponds to a correction of order
$|f-1|\simeq (1-q^2)/5 \simeq 16\%$ for $q=0.7$.  Making HST14176+5226 
a $q=0.7$ galaxy viewed pole on leads to corrections large enough to make this system consistent with 
a nearly flat rather than a rising rotation curve, thereby eliminating the apparent
disagreement between this lens and the results of the statistical analysis.  
Future analyses of lens dispersions need to treat these systematic
uncertainties more carefully since for many purposes they are already larger
than the formal measurement errors.

\section{Time Delay Measurements} \label{sec:delays}

The Party line these days is that the Hubble constant is a known quantity
based either on local estimates ($H_0=72\pm8$~km/s/Mpc, Freedman et al.~\cite{Freedman2001p47})
or models of CMB fluctuations assuming a flat universe with a cosmological constant
($H_0=72\pm5$~km/s/Mpc, Spergel et al.~\cite{Spergel2003p175}).  For the purposes
of this review, let us assume that the Party is correct and consider its implications.
Gravitational lens time delays measure a combination of the Hubble constant and the
surface density $\langle\kappa_{AB}\rangle$ in the annulus between the images for which the 
delay was measured, so if we know $H_0$ then time delays provide a very simple
means of breaking the degeneracy in Eqn.~\ref{eqn:onezone}.  As in 
\S\ref{sec:lens}, we will discuss only the effects of the radial density distribution.
-- the effects of the angular distribution are well understood and are reviewed in 
Kochanek~(\cite{Kochanek2004a}).    

The key to understanding time delays comes from  Gorenstein et al.
(\cite{Gorenstein1988p693}, Kochanek \cite{Kochanek2002p25},
see also Saha~\cite{Saha2000p1654}) who showed that the time
delay in a circular lens depends only on the image positions and {\it the surface density
$\kappa(R)$ in the annulus between the images} (Fig.~\ref{fig:geometry}). 
A useful approximation is to assume that the surface density
in the annulus can be {\it locally} approximated by a power law,
$\kappa(R)\propto R^{1-n}$ for $R_B<R<R_A$,
with a mean surface density in the annulus of $\langle\kappa_{AB}\rangle$.
The time delay between the images is then (Kochanek~\cite{Kochanek2002p25})
\begin{equation}
    \Delta t = 2 \Delta t_{SIS} \left[ 1-\langle\kappa_{AB}\rangle - { 1- n\langle\kappa_{AB}\rangle \over 12 }
           \left( { \dr \over \rbar } \right)^2 +
         O\left(\left( { \dr \over \rbar } \right)^4 \right) \right]
  \label{eqn:delay1}
\end{equation}
where 
\begin{equation}
  \Delta t_{SIS} = { 1 \over 2 } { D_{OL} D_{OS} \over c D_{LS} }
             \left( R_A^2-R_B^2 \right) \propto H_0^{-1}
  \label{eqn:delaysis}
\end{equation}
is the time delay for a singular isothermal sphere with comoving angular diameter
distances $D_{OL}$, $D_{OS}$ and $D_{LS}$ between the observer, lens and 
source.\footnote{
  Note that with comoving angular diameter distances the extra factor of $1+z_l$
  you must keep track of when using angular diameter distances has vanished.  Almost
  all lensing calculations simplify if comoving angular diameter distances are used rather
  than ``normal'' angular diameter distances, because it is no longer necessary to keep track
  of the $1+z_l$ factors in many standard expressions.}  
The key point is that the time delay is largely determined by the average surface density
$\langle\kappa_{AB}\rangle$ in the annulus with only modest corrections from the local shape of the surface
density distribution even when $\dr/\rbar \sim 1$.  The second order
expansion is exact for a singular isothermal lens ($\langle\kappa_{AB}\rangle=1/2$, $n=2$), 
and it reproduces the time delay of a point mass lens ($\langle\kappa_{AB}\rangle=0$) to better than 1\%
even when $\dr/\rbar=1$.  

\begin{figure}[t]
\begin{center}
\centerline{\includegraphics[width=4.0in]{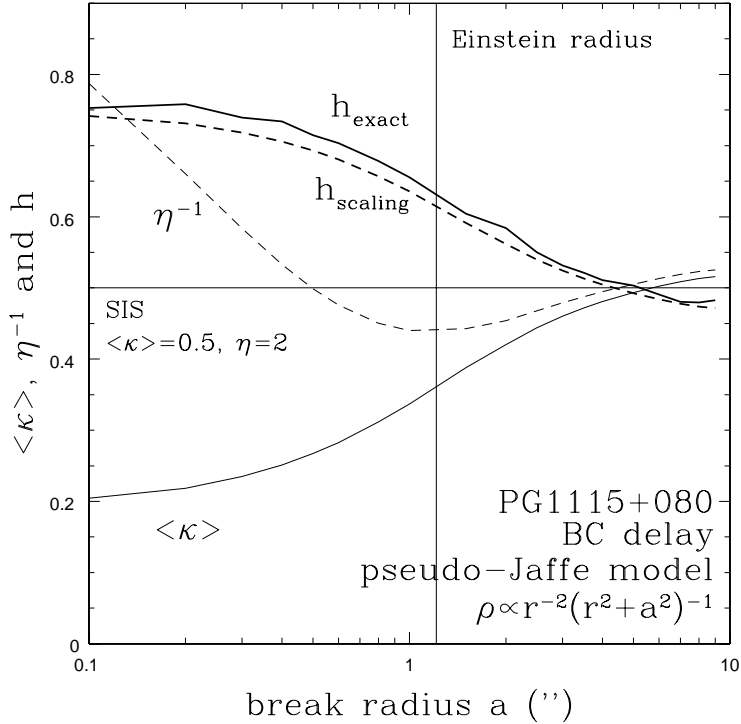}}
\end{center}
\caption{
$H_0$ estimates for PG1115+080.  The lens galaxy is modeled as an ellipsoidal
pseudo-Jaffe model, $\rho \propto r^{-2}(r^2+a^2)^{-1}$, and the nearby group
is modeled as an SIS.  As the break radius $a \rightarrow \infty$ the
pseudo-Jaffe model becomes an SIS model, and as the break radius
$a\rightarrow 0$ it becomes a point mass.  The heavy solid curve ($h_{exact}$)
shows the dependence of $H_0$ on the break radius for the exact, nonlinear
fits of the model to the PG1115+080 data.  The heavy dashed curve
($h_{scaling}$) is the value found using our simple theory
of time delays based on the surface density.  The agreement of the exact and
scaling solutions is typical. The light solid line shows the average surface
density $\langle\kappa\rangle$ in the annulus between the images, and the
light dashed line shows the {\it inverse} of the logarithmic slope $\eta$ in
the annulus ($\kappa\propto R^{1-\eta}$).
For an SIS model we would have $\langle\kappa\rangle=1/2$ and
$\eta^{-1}=1/2$, as shown by the horizontal line.  When the break radius is
large compared to the Einstein radius (indicated by the vertical line), the
surface density is slightly higher and the slope is slightly shallower than
for the SIS model because of the added surface density from the group.  As we
make the lens galaxy more compact by reducing the break radius, the surface
density decreases and the slope becomes steeper, leading to a rise in $H_0$.
As the galaxy becomes very compact, the surface density near the Einstein ring
is dominated by the group rather than the galaxy, so the surface density
approaches a constant and the logarithmic slope approaches the value
corresponding to a constant density sheet ($\eta=1$).
\label{fig:pg1115}}
\end{figure}

\begin{figure}[t]
\begin{center}
\centerline{ \includegraphics[width=4.0in]{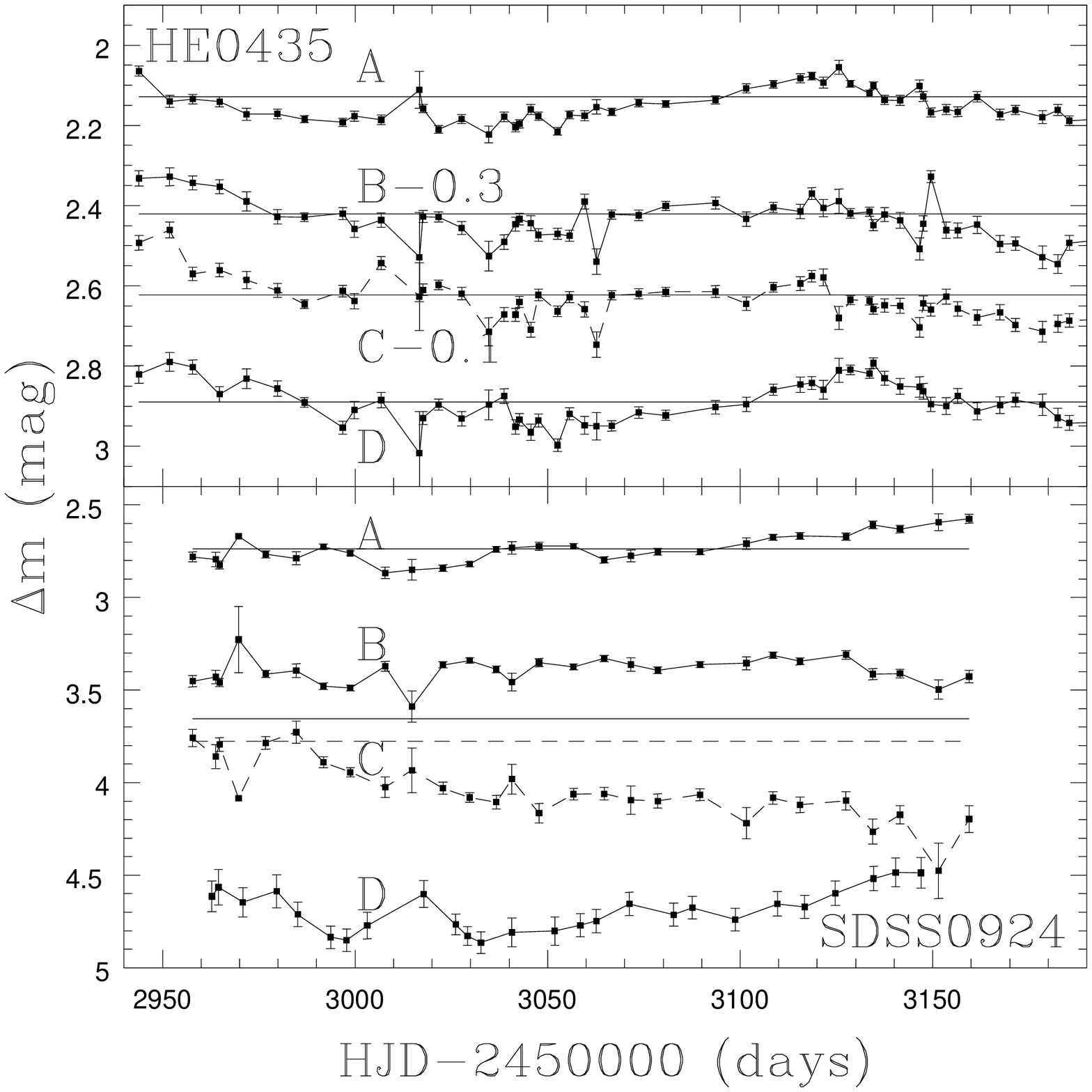} }
\end{center}
\caption{
  Our raw light curves for HE0435--1223 (top) and SDSS0924+0219 (bottom).  
  The HE0435--1223 light curve was shifted by 80 days.
   We have measured
  the longest time delay in the system (A to D) to 10\% accuracy from this data.  The
  B and C images have shorter delays -- their different light curve shapes are due to
  an additional steady fading created by microlensing.  In the current season we already
  see additional intrinsic variability that is increasing the accuracy of the time
  delay measurements.  For SDSS0924+0219 we see
  little evidence for intrinsic variability -- the changes in the flux ratios are
  entirely due to microlensing.    The horizontal lines, referenced to the mean of the
  A light curve, show the flux ratios when the lens was discovered (HJD--2450000=2259).
  In the first images of the new season, the fluxes of images A, C and D are little
  changed, but image B is 0.5 mag fainter.  Eventually we expect a dramatic change
  as image D returns to its expected flux (roughly that of A).
   }
\label{fig:lcurves}
\end{figure}

This relationship between annular surface density and time delays is the 
fundamental relationship to remember.  A popular parametric example
is the scaling of the time delay for global power law models with 
$\kappa \propto R^{1-n}$.  In these models the mass density scales
as $\rho \propto r^{-n}$ and circular velocities or ray deflections
scale as $v_c^2 \propto \alpha \propto R^{2-n}$ where $n=2$ is
the singular isothermal (SIS) profile producing a flat rotation curve
or a constant deflection.  In these models the time delay
is $\Delta t = \Delta t_{SIS} (n-1)$ to lowest order in $\dr/\rbar$
(Witt et al.~\cite{Witt2000p98}).  Although the logarithmic slope
$n$ is a global parameter of the model, it enters the time delay
only through the general requirements of Eqn.~\ref{eqn:onezone}
and the surface density of the model near the Einstein ring -- 
the surface density near the Einstein ring is $\langle\kappa_{AB}\rangle=(3-n)/2$, 
so $1-\langle\kappa_{AB}\rangle=(n-1)$ explains the scaling of the time delay.
Fig.~\ref{fig:pg1115} shows a more complicated example where we
compare estimates of $H_0$ from a complex parametric model for 
PG1115+080 consisting of an ellipsoidal lens plus its parent 
group to the value we would get simply calculating the surface
density of the model and using the generalization of 
Eqn.~\ref{eqn:delay1} for the angular structure of the lens
(Kochanek~\cite{Kochanek2002p25}).
As we would expect from our simple theory, the surface density of
the lens and its local slope are all the information needed to
predict the scaling of the Hubble constant with the model
parameters.

There are presently twelve lenses with published time delay measurements of
varying accuracy:
B0218+357 (Biggs et al.~\cite{Biggs1999p349}),
RXJ0911+0551 (Hjorth et al.~\cite{Hjorth2002p11}),
FBQ0951+2635 (Jakobsson et al.~\cite{Jakobsson2004p1}),
Q0957+561 (Kundi\'c et al.~\cite{Kundic1997p75}),
HE1104--1805 (Ofek \& Maoz~\cite{Ofek2003p101}),
PG1115+080 (Schechter et al.~\cite{Schechter1997p85}),
B1422+231 (Patnaik \& Narasimha~\cite{Patnaik2001p1403}),
SBS1520+530 (Burud et al.~\cite{Burud2002p481}),
B1600+434 (Burud et al.~\cite{Burud2000p117}, Koopmans et al.~\cite{Koopmans2000p391}),
B1608+656 (Fassnacht et al.~\cite{Fassnacht2002p823}),
PKS1830--211 (Lovell et al.~\cite{Lovell1998p51}) and
HE2149--2745 (Burud et al.~\cite{Burud2002p71}).
The interpretation of these delays is controversial
(e.g. Williams \& Saha~\cite{Williams2000p439}, Kochanek~\cite{Kochanek2002p25}, 
Koopmans et al.~\cite{Koopmans2003p70}
and references therein), but the problem lies mainly with the limitations
of the existing data.  The first problem is that several of the lenses
are poorly suited to analysis because their photometric structure makes
it difficult to perform the accurate astrometry of the images relative
to the lens galaxy needed to interpret the delays for any
mass distribution.  For example, the quasar images are too bright in
B0218+357, the lens is too faint in PKS1830--211, and the position
estimates depend on accurate corrections for extinction in B1600+434
and B1608+656.   The second problem is that many of the delays are
not known with sufficient accuracy because the monitoring programs
were terminated after the initial delay estimate.  We need time delay
measurement accuracies of 5\% or better in order to have
uncertainties in the surface density $\langle\kappa_{AB}\rangle$ that are small compared to the
regime of interest ($0.1 \ltorder \langle\kappa_{AB}\rangle \ltorder 0.5$, see Fig.~1).
Half of the lenses (FBQ0951+2635, HE1104--1805, PG1115+080, B1422+231, PKS1830--211 
and HE2149--2745) have unacceptably high delay uncertainties.  The third problem
is that we have too few systems to adequately understand environmental effects.
While Keeton \& Zabludoff~(\cite{Keeton2004p660}) somewhat exaggerate the 
importance of environmental effects on the interpretation of time delays 
(much of the spread in their distributions is due to the uncertainties in
the time delay measurement rather than the effects of the group), it is
certainly true that environments affect the interpretation because the parent
group or cluster of the lens galaxy contributes to $\langle\kappa_{AB}\rangle$.
For example, B1608+656 has two interacting lens galaxies inside the Einstein
radius and the lens galaxies of RXJ0911+0551 and Q0957+561 are members of
relatively massive clusters where
the cluster potential is important to interpreting the time delays.
In fact, if we assume the two time delay lenses in clusters (Q0957+561 and RXJ0911+0551)
have similar halo structures to the other lenses, then in order to obtain the same value
of $H_0$ as that of the isolated lenses, the clusters must be
contributing significantly to the average surface density
($\kappa_{cluster} \sim 0.1$--$0.3$).
Thus, by the time the sample is narrowed to isolated, early-type lenses
with good astrometry, good photometry and accurate time delays, almost nothing
is left.

As a result, I find much of the recent debate in lensing circles about time delay lenses
to be sterile.  That one person's ``Golden Lens'' can match the Party's value of
$H_0$ with a flat rotation curve is largely meaningless when someone else's
``Golden Lens'' requires a constant mass-to-light ratio to do so unless you
can also provide a coherent explanation of why the mass distributions of the
two lenses should be different.  Hence the philosophy we adopt here -- the Party
says $H_0$ is known and should not be challenged unless you want to count
trees in Siberia, so we should focus on why we cannot produce a coherent 
story about the structure of lens galaxies at the Party's value of $H_0$ before
deciding to attempt a revolution.   One possibility, that many delay measurements
are less accurate than believed, is easily solved by monitoring the lenses
longer.  The other possibility, that the halo structures of early-type
galaxies are heterogeneous rather than homogeneous, can be addressed 
independent of the actual value of $H_0$.

We can make some initial observations about the homogeneity of halos.  If we take the
four relatively isolated bulge-dominated time delay lenses 
(PG1115+080, SBS1520+530, B1600+434 and HE2149--2745), then we can
show that for the same value of $H_0=100 h$\hunits, the lenses must
have similar mass distributions.  For $H_0 \simeq 50$\hunits\ they
can have flat rotation curves, while for $H_0 \simeq 72$\hunits\
their mass distributions must be nearly constant $M/L$ (Kochanek~\cite{Kochanek2002p1}). 
While we cannot (without knowing $H_0$) use the time delays to
determine which mass distribution is correct, we know that their
mass distributions are homogeneous.  If we use our surface density
formalism, they have mean surface densities of
$$
  \langle\kappa_{AB}\rangle = 1.14 - 1.21 h \pm 0.04,
$$
with an upper limit of $\sigma_\kappa<0.07$ on the intrinsic scatter in $\langle\kappa_{AB}\rangle$
between the halos.  The uncertainties are dominated by the time delay measurement
uncertainties for the lenses (Kochanek~\cite{Kochanek2002p25}).  This result
is somewhat puzzling when we compare it to the stellar dynamical results.   
Treu \& Koopmans~(\cite{Treu2002p6}) argue that PG1115+080 must have a 
falling rotation curve while the typical lens they have studied has a 
slightly rising rotation curve (Treu \& Koopmans~\cite{Treu2004p739}).  
But the homogeneity of the surface density estimates based on the time delays
indicates that if PG1115+080 has a falling rotation curve then SBS1520+530,
B1600+434 and HE2149--2745 must as well.  It is odd that the typical time
delay lens would have a falling rotation curve while the typical stellar
dynamical lens would have a rising rotation curve -- the most likely 
solution is a systematic bias in one of the methods.  This will be easily
addressed once we have a larger overlapping sample of galaxies with time
delay and velocity dispersion measurements.

\begin{figure}[t]
\centerline{
   \includegraphics[width=2.7in]{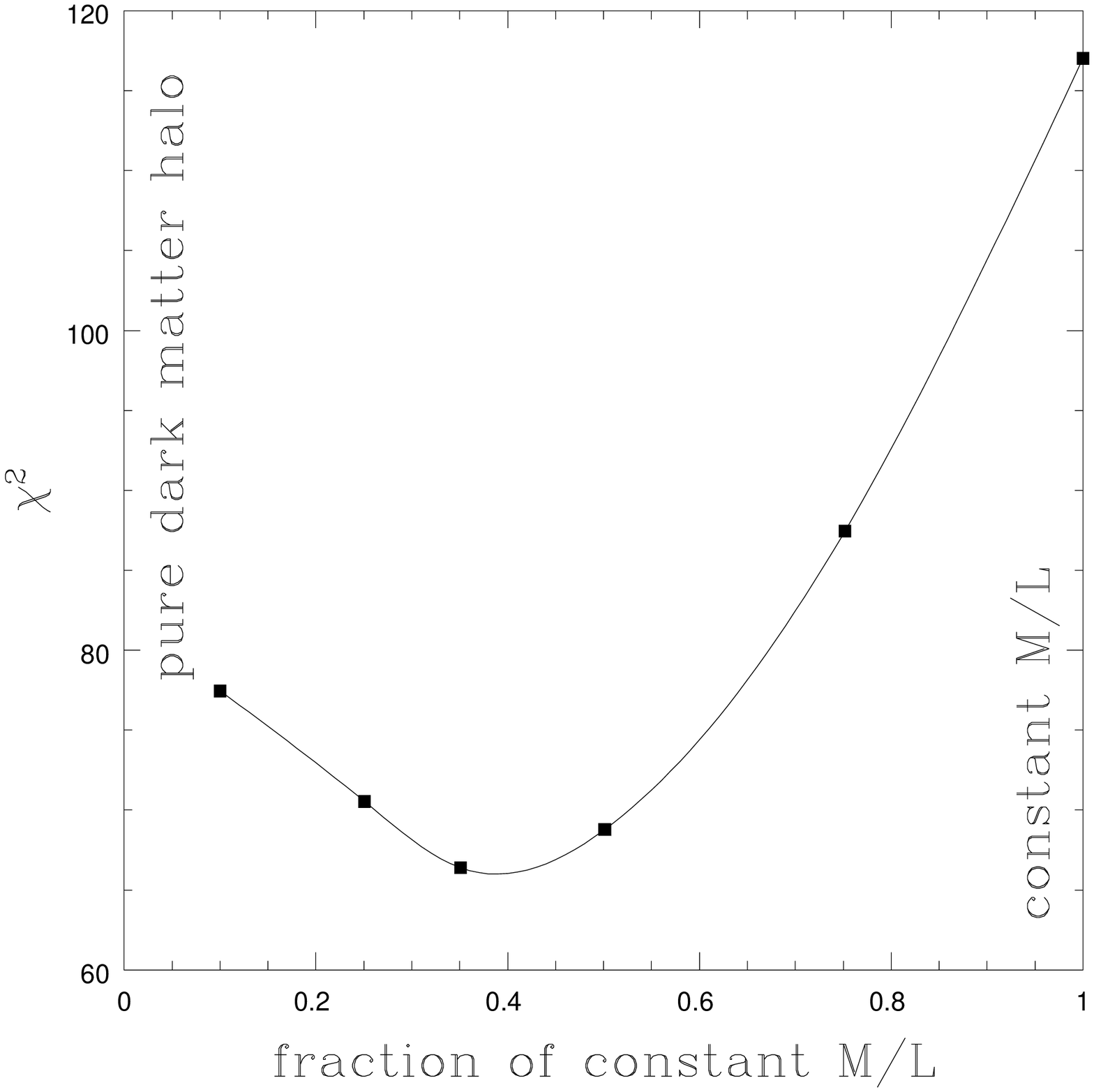}
   \includegraphics[width=3.0in]{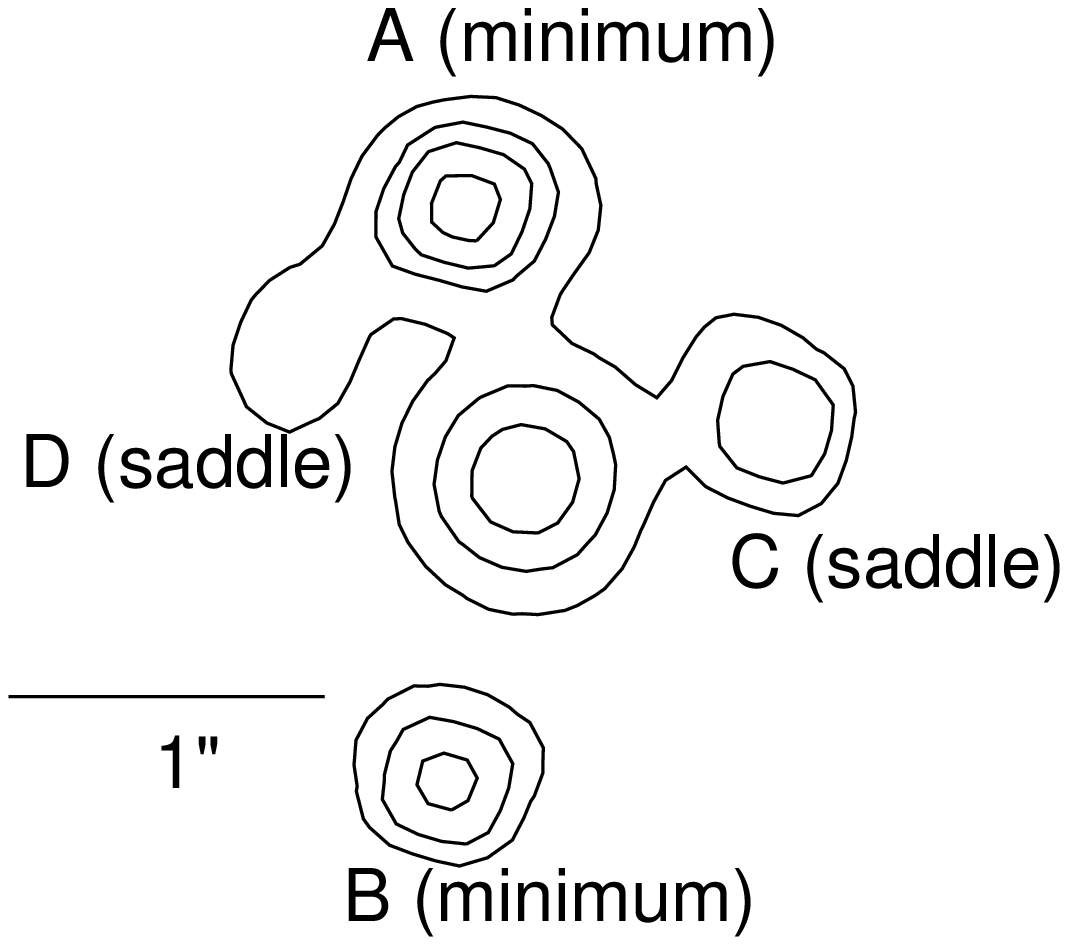}
   }
\caption { (LEFT)
Dark matter in HE0435--1223.  We fit the constraints, including the time delay,
starting with a constant $M/L$ de Vaucouleurs model for the stars and then adding
an NFW halo (similar to Fig.~1).  Here we show the goodness of fit statistic $\chi^2$
as a function of the mass of the de Vaucouleurs model relative to its mass in the
constant $M/L$ model.  The best fit is that the mass of the stellar component can
be only $(39\pm6)\%$ that of a constant $M/L$ model.  In this regime, the overall
mass distribution has a rotation curve that is very close to flat.  At the minimum
we have a good fit to the constraints including the geometry of the Einstein ring
of the quasar host galaxy.
  }
 \label{fig:dm0435}
\caption{ (RIGHT)
  An HST image of SDSS0924+0219 stretched to show the flux ratios of the quasar images A--D.
  In the absence of microlensing (or substructure), the flux of image D should be almost
  equal to that of image A.  Instead, it is an order of magnitude fainter.  If this
  flux ratio anomaly is due to microlensing, we should see it return to its normal flux
  with quite dramatic color changes for probing the structure of the accretion disk.
  }
 \label{fig:sdss0924}
\end{figure}

\begin{figure}[t]
\centerline{
   \includegraphics[width=2.5in]{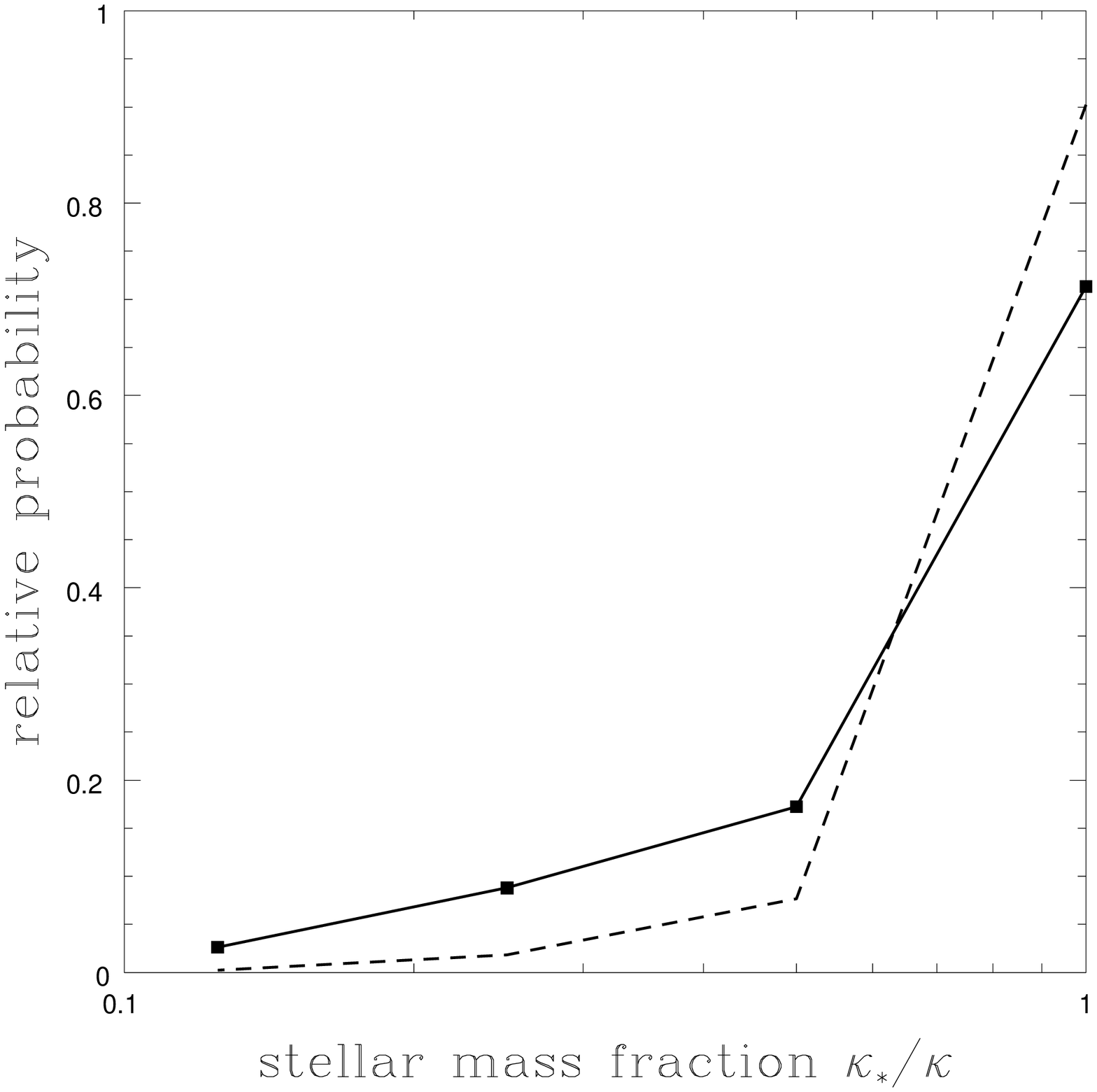}
   \includegraphics[width=2.5in]{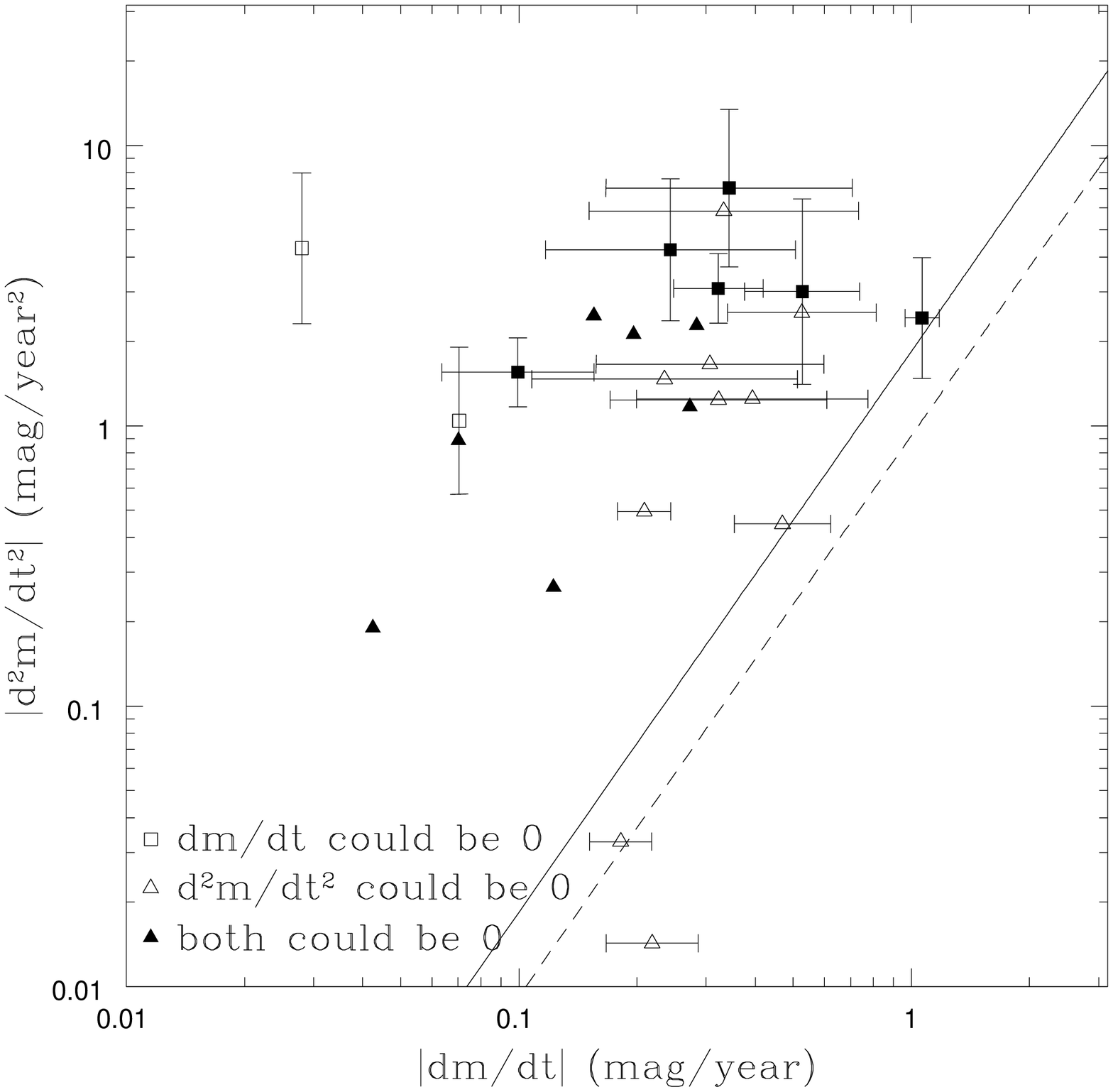}
   }
\caption{ (LEFT)
  The stellar mass fraction $\kappa_*/\kappa$ in Q2237+0305.  The dashed curve shows the effect
  of restricting the microlens mass range to $0.2h^2M_\odot < \langle M \rangle < 2h^2 M_\odot$.
  This eliminates solutions with high effective velocities,
  so higher stellar densities are needed to produce the same level of variability.
  We see the Q2237+0305 images through the bulge of a nearby spiral galaxy, so we
  expect $\kappa_*/\kappa \simeq 1$ for this lens.
  }
 \label{fig:kstar}
\caption{ (RIGHT)
  Microlensing rates for the first season.  The points show fits to the first $dm/dt$ and
  second $d^2m/dt^2$ derivatives of the delay-corrected differential light curves for
  the lensed images in our monitoring sample with adequate data in mid-2004.  The point-type changes
  depending on whether the derivative measurements are consistent with zero.  The solid
  line shows the relation expected for a point source crossing a fold caustic, and the
  dashed line shows the lower limit on $d^2m/dt^2$ for a point mass lens.  The second
  derivatives are not very reliable at present because many of the light curves used 
  for the estimates were quite short (months) -- we do not see multi-magnitude 
  variations in any of our light curves.
  }
 \label{fig:micro}
\end{figure}

At this point I believe the only good solution is to simply measure many more
delays with significantly higher accuracies.  With 25 accurate ($<5\%$) time
delays in systems with good ancillary data (photometry, astrometry), it will
be relatively simple to determine the heterogeneity or homogeneity of early-type
galaxy halos.  The delay sample will also better overlap systems with extra
constraints or stellar dynamical measurements.  
For a fixed value of $H_0$, time delay measurements are a
significantly better probe of the halo structure than either statistical
studies or stellar velocity dispersion measurements because they are both
more accurate and directly measure the quantity of greatest interest, the
surface density in the transition region from luminous to dark matter,
$\langle\kappa_{AB}\rangle$.  We have started such a program
using the SMARTS 1.3m telescope in the South and an array of telescopes in the
North (MDM, FLWO, APO $\cdots$) to monitor roughly 25 lenses 1-2 times/week
when they are visible.  Since starting in early 2004 we have measured two
new time delays in HE0435--1223 and SDSS1004+4112.  We also see an abundance
of microlensing variability, whose utility we discuss in \S\ref{sec:micro}.
Fig.~\ref{fig:lcurves} shows the light curves 
for HE0435--1223 and SDSS0924+0129    
from the first season of monitoring. 

Fig.~\ref{fig:dm0435} shows the results of fitting a model consisting of the luminous
lens galaxy and an NFW dark matter halo to HE0435--1223 constrained by the
time delay measurement with the Hubble constant fixed to $H_0=72$\hunits.
If we start from a constant $M/L$ model and then reduce the mass of the 
luminous galaxy while adding the NFW halo, we see the goodness of fit
improves with a minimum where the luminous galaxy has only $(39\pm6)\%$
of its mass in the constant $M/L$ model.  As we improve the delay from
its current 10\% accuracy, these uncertainties will shrink more or 
less proportionately.  

The other obvious result from our first season is that microlensing of 
lensed quasars is ubiquitous.  In Fig.~\ref{fig:lcurves} you can see that
the light curves of images A and D in HE0435--1223 are nearly identical
while those for B and C look different.  Images A and D have the longest
delay and are the pair for which we have a time delay measurement. In
the first season images B and C were both fading by $0.18\pm 0.02$~mag/year
relative to A and D due to microlensing.  Other light curves, like
the ones for SDSS0924+0129 shown in Fig.~\ref{fig:lcurves}, show only
brightness variations due to microlensing.  Collecting and analyzing
the microlensing variability is important because it provides our
final probe of halo structure using lenses. 

\section{Microlensing} \label{sec:micro}

Microlensing variability depends on both the surface density in stars $\kappa_*$
and the total surface density $\kappa$.  The regime for a standard halo model
is that the total surface density is still relatively high ($\kappa \sim 0.5$)
but the stars make up a small fraction of the total ($\kappa_*/\kappa \sim 0.1$,
see Fig.~\ref{fig:halo}).
In this regime, microlensing (or substructure for that matter) has a distinctive
statistical pattern that strongly distinguishes between images that are saddle 
points and minima (Schechter \& Wambsganss~\cite{Schechter2002p685}).  In particular,
there is considerable phase space for significantly demagnifying saddle point
images.  The most dramatic example of this is the (saddle point) image D of 
SDSS0924+0219, which is almost ten times fainter than expected (see Fig.~\ref{fig:sdss0924}).
We can already see in the SDSS0924+0219 light curves of Fig.~\ref{fig:lcurves}
that the images are being microlensed, but if the suppression of image D is
due to microlensing we should eventually see it brighten dramatically, overshooting
the flux of image A.  

Because the statistical properties of the microlensing patterns depend on 
$\kappa$ and $\kappa_*/\kappa$ we can constrain them by analyzing the
light curves.  In Kochanek~(\cite{Kochanek2004p58}) we developed a method
for analyzing microlensing light curves and applied it to the only (hitherto)
well-monitored lens Q2237+0305, focusing on the OGLE light curves of
Wozniak et al.~(\cite{Wozniak2000p65}).  In this system we see four images
of a quasar buried in the bulge of a low redshift spiral galaxy so we
would expect to find that $\kappa \simeq \kappa_*$, and this is borne out
when we analyze the microlensing variability as shown in Fig.~\ref{fig:kstar}.  
Even after including all the other uncertainties (sources sizes, relative
velocities $\cdots$) we find that $\kappa_*/\kappa > 0.5$ if we include
no prior assumption on the mean stellar masses, and $\kappa_*/\kappa > 0.7$ if we
force the mean stellar mass to be in the range of $0.1$-$1.0M_\odot$.  

We do not yet have long enough light curves to make a similar analysis
of any other lens, but we are beginning to get there.  Fig.~\ref{fig:micro}
shows a crude view of the amount of microlensing, where we estimate
the first and second derivatives of the light curves due to microlensing
from the data available in mid-2004.  Most images have measurable first
and second derivatives that cannot be attributed to intrinsic variability.  
Since we want
accurate time delays rather than simply quitting once the delay is
measured to 10\%, we will be monitoring systems like HE0435--1223 for
many more years.  The same longer time baselines are needed
to obtain useful constraints on the statistical properties of the
magnification pattern produced by the stars.   The microlensing
variability can also be used to probe the structure of quasar accretion
disks.  For example, even the limited data for SDSS0924+0219 is
sufficient to show that it must have an accretion disk that
is significantly smaller than that in Q2237+0305, 
as we might expect from the lower luminosity of SDSS0924+0219.

\section{Conclusions}\label{sec:concl}

The overall summary of this review is that the future is bright, but 
the present is painful.  
On the positive side, we understand the physics of constraining
mass distributions with lenses very well.  In general, analyses based on 
image constraints, statistics, velocity dispersions, time delays and 
microlensing are all generally consistent and indicate that early-type
galaxies have appreciable dark matter on intermediate scales ($r \sim 2R_e$)
leading to a net mass distribution that is close to isothermal
for the typical lens. 
On the negative side, we have not communicated our understanding of how
gravitational lenses constrain mass distributions very well.  Arguments
about methods (parametric versus non-parametric, statistics versus time
delays versus dispersions and so on) tend to obscure rather than 
illuminate the underlying unity of the results.

The fundamental question at this point is the degree to which early-type
galaxy halos are heterogeneous in their structures on these scales.  Most
of this is driven by a fairly familiar astronomical problem.  First, there
is a preferred, average, naive, straw man, or whatever mass model in which
early-type galaxies follow the conspiracy of late-type galaxies and have
nearly flat rotation curves with the contribution from the dark matter
kicking in just as the contribution from the luminous matter fades.  
Theoretical halos with the expected baryonic mass fraction would be
expected to lie close to this model, usually with a slowly falling
rather than a truly flat rotation curve.  Second, when we apply the available
methods to the lens sample, we find that this is just about right, but
there are signs that there is appreciable scatter about the mean. 
Because of small sample sizes, measurement errors and systematic errors
we cannot yet be sure whether this scatter is real and the halo 
structure is genuinely heterogeneous or if we have simply underestimated
our measurement uncertainties.  This puts the field in a confused state where 
everyone argues for their object or method but cannot produce a synthesis
of all the available results to produce a convincing final answer. 
Until we can achieve this synthesis we will continue to have a problem.

Fortunately the way out is straightforward -- more data.  No one has
found a fundamental flaw in any of the methods.  The problem is simply 
that the samples are too small and insufficiently overlapping to produce a
clear result.  My own feeling is that the method with the least
systematic problems for addressing the homogeneity or heterogeneity
of early-type galaxy halos is monitoring lenses to measure accurate
time delays and comparing the structure of lens galaxies at a fixed
value of $H_0$.  For fixed $H_0$, time delays measure the surface
density of the galaxy near the lensed images essentially to the
accuracy of the time delay measurement, and the surface density
near the lensed images is exactly the quantity needed to understand
the halo structure.  Moreover, the long term monitoring needed to
measure accurate time delays also provides microlensing variability
light curves that can be used to constrain the fraction of the 
surface density in the form of stars.  There is a very convenient
coincidence that the part of parameter space that should be
occupied by the lens galaxies in a standard CDM model also has unique 
microlensing characteristics.

\begin{acknowledgments}
I thank C. Morgan, D. Rusin and J. Winn for their comments.
The author is supported  by the NASA ATP grant NAG5-9265 and
by grants HST-GO-8804 and 9744 from the Space Telescope
Science Institute. 
\end{acknowledgments}

\end{document}